\documentclass[reprint,
superscriptaddress,
altaffilletter,
amsmath,amssymb,
 aps,
]{revtex4-1}

\usepackage{graphicx}
\usepackage{dcolumn}
\usepackage{bm}
\usepackage{stackengine}
\usepackage{textcomp}
\usepackage[hidelinks]{hyperref}
\usepackage[nameinlink,capitalise]{cleveref}
\usepackage[usenames, dvipsnames]{color}
\usepackage{float}

\usepackage{color}

\begin{document}
\preprint{APS/123-QED}

\title{Measurement of the scintillation and ionization response of liquid xenon at MeV energies in the EXO-200 experiment}

\author{G.~Anton}\affiliation{Erlangen Centre for Astroparticle Physics (ECAP), Friedrich-Alexander-University Erlangen-N\"urnberg, Erlangen 91058, Germany}
\author{I.~Badhrees}\altaffiliation{Permanent position with King Abdulaziz City for Science and Technology, Riyadh, Saudi Arabia}\affiliation{Physics Department, Carleton University, Ottawa, Ontario K1S 5B6, Canada}
\author{P.S.~Barbeau}\affiliation{Department of Physics, Duke University, and Triangle Universities Nuclear Laboratory (TUNL), Durham, North Carolina 27708, USA}
\author{D.~Beck}\affiliation{Physics Department, University of Illinois, Urbana-Champaign, Illinois 61801, USA}
\author{V.~Belov}\affiliation{Institute for Theoretical and Experimental Physics named by A.I. Alikhanov of National Research Centre ``Kurchatov Institute'', 117218, Moscow, Russia}
\author{T.~Bhatta}\affiliation{Department of Physics, University of South Dakota, Vermillion, South Dakota 57069, USA}
\author{M.~Breidenbach}\affiliation{SLAC National Accelerator Laboratory, Menlo Park, California 94025, USA}
\author{T.~Brunner}\affiliation{Physics Department, McGill University, Montr\'eal, Qu\'ebec H3A 2T8, Canada}\affiliation{TRIUMF, Vancouver, British Columbia V6T 2A3, Canada}
\author{G.F.~Cao}\affiliation{Institute of High Energy Physics, Beijing, China}
\author{W.R.~Cen}\affiliation{Institute of High Energy Physics, Beijing, China}
% ...

\date{\today}
\pacs{Valid PACS appear here}
~\begin{abstract}
Liquid xenon (LXe) is employed in a number of current and future detectors for rare event searches. We use the EXO-200 experimental data to measure the absolute scintillation and ionization yields generated by $\gamma$ interactions from $^{228}$Th (2615~keV), $^{226}$Ra (1764~keV) and $^{60}$Co (1332~keV and 1173~keV) calibration sources, over a range of electric fields. The $W$-value that defines the recombination-independent energy scale is measured to be $11.5~\pm~0.5$~(syst.)~$\pm~0.1$~(stat.) eV. These data are also used to measure the recombination fluctuations in the number of electrons and photons produced by the calibration sources at the MeV-scale, which deviate from extrapolations of lower-energy data. Additionally, a semi-empirical model for the energy resolution of the detector is developed, which is used to constrain the recombination efficiency, i.e., the fraction of recombined electrons that result in the emission of a detectable photon. Detailed measurements of the absolute charge and light yields for MeV-scale electron recoils are important for predicting the performance of future neutrinoless double beta decay detectors.
\end{abstract}
\maketitle
\section{\label{intro}Introduction}
The EXO-200 experiment searched for neutrinoless double beta decay ($0\nu\beta\beta$) in $^{136}$Xe using liquid xenon (LXe) as both the source of the decay and the detector medium~\cite{EXOlatest}. Using this technique, EXO-200 set a lower limit on the half-life of $0\nu\beta\beta$ of $T_{1/2}^{0\nu\beta\beta} > 3.5\times10^{25}$ years~\cite{EXO-200_final,EXOlatest}.  Future detectors employing LXe, such as nEXO, are planned to reach half-life sensitivities approaching 10$^{28}$~years~\cite{Albert:2017hjq,Kharusi:2018eqi}. Accurately measuring the response of LXe to MeV-scale electron recoils is directly relevant to understanding the ionization and scintillation process and its impact on the energy resolution  for nEXO~\cite{Albert:2017hjq} and other future $0\nu\beta\beta$ detectors~\cite{Akerib:2019dgs,Agostini:2020adk}.

This paper describes the first absolutely calibrated measurement of the response of liquid xenon in the range of energies that are directly relevant to predicting the sensitivity of future neutrinoless double beta decay (0vbb) detectors. 

Particles interacting in LXe can deposit a portion of their energy as scintillation and ionization, which can be detected by incorporating the LXe into a time projection chamber (TPC).  
In the ionization process, particles such as $\gamma$-ray photons, $\alpha$ particles, or energetic electrons deposit their energy in LXe through different physical mechanisms and produce a number of electron-ion pairs and excited xenon atoms (``excitons'')~\cite{Aprile}, as well as some loss of energy to undetectable channels (e.g. atomic motion or ``heat''). In the scintillation process, two routes are possible: either direct excitation of Xe atoms or electron-ion recombination. Both processes lead to the production of excitons which form excited dimers, Xe$_2^*$, and then de-excite with the emission of a $\sim$178~nm vacuum ultraviolet (VUV) photon~\cite{Aprile}. The relative number of electrons and photons collected from an event is anti-correlated, as first measured in~\cite{PhysRevB.68.054201}, and depends on the electric field applied across the TPC.  As the electric field increases, more ionized electrons can be drifted away from the interaction site, reducing the number of photons produced through recombination. 

In this paper we present a measurement of absolute light and charge yields in LXe, from which we can extract the ``$W$-value,'' which is defined as the average energy needed to produce a quantum of either charge or light (i.e., either an electron or VUV photon). This definition of W-value follows the same approach in Ref.~\cite{NEST_2011,Akerib,MiX,Dahl,Shutt:2006ed,PhysRevD.97.102008,Xen1T}, and is to be distinguished from notations used in older literature such as Ref.~\cite{Doke,Chepel:2012sj}, where ``$W$" represents the average energy required to produce an electron-ion pair. In this work the ionization-only value from Ref.~\cite{Doke,Chepel:2012sj} is denoted as ``$W_i$" instead. LXe detectors are widely used in dark matter searches and previous measurements of the light and charge yields using those detectors have primarily focused on low energy electron and nuclear recoils ($\lesssim 500$~keV)~\cite{NEST_2011}.  The measurements described here are focused on the detector response of LXe to higher energy ($\sim$1-2.5~MeV) electron recoils.  

In Sec.~\ref{prev_meas}, a summary is presented of previous measurements of the absolute charge and light yields in LXe and the corresponding $W$-values.  Sec.~\ref{setup} provides a brief overview of the EXO-200 detector and its charge and light sensors, while the yield measurement and its comparison to the Noble Element Simulation Technique (NEST)~\cite{NEST_2011} is shown in Sec.~\ref{yield}. In Sec.~\ref{rotationangle}, the energy resolution for $\gamma$ events at different electric fields is measured and compared to the values predicted by a semi-empirical model. An important input to the resolution model---the recombination fluctuations at a variety of energies above 1~MeV---are measured and compared to previous lower-energy data. Finally, this section presents constraints on the recombination efficiency from the comparison of the resolution model to data. 

\section{\label{prev_meas}Previous measurements}
A number of previous detectors have measured the charge and light response of LXe to $\alpha$-, $\beta$-, and $\gamma$-induced electron recoils, and neutron-induced nuclear recoils (see, e.g.,~\cite{Akerib,Hornet,TANAKA2001454,Akimov:2014cha,PhysRevLett.97.081302,Doke,Manalaysay:2009yq,Aprile:2012an,Baudis:2017xov,Aprile:2017xxh,MiX,PhysRevD.88.012006,PhysRevD.72.072006,PhysRevC.84.045805}).  In addition, the Noble Element Simulation Technique (NEST) software tool has been developed to provide an empirical model to simulate the charge and light responses for LXe under different electric fields and for the various particle types~\cite{NEST_2011,szydagis_m_2018_1314669}.

In general, the number of detected electrons and photons depends on the electric field applied to the LXe, since a field-dependent fraction, $r$, of the initially produced electron-ion pairs can recombine to produce excitons that then emit a photon.  For an initial population of $n_i$ electron-ion pairs and $n_{ex}$ excitons, the maximum number of detectable electrons, $n_q$, at a given field is $n_q = (1-r) n_i$ and the number of detectable photons is $n_p = n_{ex} + r n_i$, under the assumption that each recombining electron-ion pair produces an exciton, which in turn produces a photon. This assumption will be relaxed in Sec.~\ref{intrinsicrho}, where constraints on the absolute recombination efficiency are studied using EXO-200 experimental data. Under the above assumption, it is possible to define a recombination-independent $W$-value, $W = E/(n_q + n_p)$ which corresponds to the energy required to produce a single detectable quantum of either type.  Since $n_q + n_p = (1-r)n_i + n_{ex} + r n_i = (1+\alpha)n_i$ for $\alpha = n_{ex}/n_i$, this definition of $W$ does not depend on electric field. {We note, as pointed out in Ref.~\cite{Dahl}, that if the efficiencies for an exciton or recombining electron-ion pair to create a detectable photon ($\epsilon_e$ or $\epsilon_r$, respectively) differ from unity, a recombination-independent energy scale can still be defined. In this case, the expression above for $W$ is unchanged, but $\alpha = \frac{\epsilon_e}{\epsilon_r}\left(\frac{n_{ex}}{n_i}\right)$ and $n_p$ would denote the number of recombining electron-ion pairs needed to produce the observed scintillation signal if there were no direct exciton production.} 

Using these definitions, we can also define the energy required to produce a single electron-ion pair prior to recombination, $W_i = E/n_i$ such that $W = W_i/(1+\alpha)$.
Since $n_q + n_p$ is constant prior to detection, if an absolute calibration of the electron count is performed (see Sec.~\ref{yield1}), the overall light detection efficiency can be calculated from the change in the detected number of electrons and photons as $r$ is varied with electric field~\cite{Shutt:2006ed,Dahl}.  This allows the light detection efficiency to be absolutely calibrated from the charge signal, which is important since the detection efficiency for charge can be nearly unity for practical detectors, while the overall efficiency for detecting VUV photons is typically only $\sim 10-20$\%~\cite{Akerib,PhysRevD.97.102008,Xen1T,PandaX,MiX} and can be difficult to model without empirical measurements.

In this work, we are primarily interested in energy depositions from MeV-scale $\beta$ and $\gamma$ particles.  A number of previous measurements of $W_i$ and $W$ exist in the literature for these particle types, with a significant spread in the reported values, as shown in Table \ref{prev-W}. For example, early measurements for 976~keV conversion electrons from a $^{207}$Bi source found $W_i = 15.6 \pm 0.3$~eV~\cite{Takahashi}, while independent measurements using 662~keV $^{137}$Cs $\gamma$s gave $W_i = 13.6 \pm 0.2$~eV~\cite{Obodovski}.  Later measurements employing an electron beam with energy per electron of 1--40~keV (and total deposited energy between 20~MeV and a few GeV) found $W_i = 9.76 \pm 0.70$~eV~\cite{Seguinot}.  More recent measurements using 122~keV $\gamma$-rays found $W_i = 16.5 \pm 0.8$~eV~\cite{Hornet}.  In Ref.~\cite{Akimov:2014cha} the authors perform a compilation of measurements of $W$ and $W_i$ and find a combined estimate of $W_i = 14.30 \pm 0.14$~eV, consistent with their data obtained from a variety of $\beta$ and $\gamma$ sources between $\sim3-700$~keV~\cite{Akimov:2014cha}.

Measurements of the recombination independent value, $W$, have also been reported.  In a reanalysis of data taken with $^{207}$Bi conversion electrons from Ref.~\cite{PhysRevB.17.2762}, Ref.~\cite{Doke} finds a value of $W = 13.8 \pm 0.9$~eV.  This value relies on their previous measurement of $W_i$~\cite{Takahashi,PhysRevB.17.2762}, which is converted to $W$ using the average of their measured value of $\alpha = 0.20$~\cite{Doke} and their calculated value of $\alpha = 0.06$~\cite{Takahashi}.  The error bar reported on $W$ indicates the difference between the measured and calculated values for $\alpha$.  Significant non-linearity is seen in the measured sum of the charge and light response, which is ascribed to possible variation in the amplifier response with rise time or loss of electrons due to electronegative impurities~\cite{Doke}.  In addition, Ref.~\cite{Shutt:2006ed} finds a value of $W = 13.46 \pm 0.29$~eV, although this value is not included in the combined average in Ref.~\cite{NEST_2011} due to a possible calibration problem~\cite{NEST_2011}. Dahl reported a value of $W = 13.7 \pm 0.2$ using 122~keV $\gamma$ events in a 30~g detector, which could be operated in single-phase or dual-phase mode~\cite{Dahl}.  The compilation in Ref.~\cite{NEST_2011} additionally references two other values: $W = 14.0$~eV, appearing only in the preprint version of Ref.~\cite{2011APh....34..679A} and excluded from the average in~\cite{NEST_2011} since it lacks an error estimate; and $W = 14.7 \pm 1.5$~eV~\cite{DOKE1990617,DOKE1993113}. The latter value comes from earlier work by a subset of the authors of Ref.~\cite{Doke} and agrees within errors with the later result.  Finally, taking the combined estimate of $W_i \sim 14.3$~eV from Ref.~\cite{Akimov:2014cha} and assuming the measured value of $\alpha \sim 0.2$~\cite{Doke,Aprile2} would correspond to $W \sim 11.9$~eV, where only an approximate value is given due to the spread in the values used to form the average and the uncertainty in the difficult to measure parameter, $\alpha$.

\begin{table}[t]
\caption{Summary of the previous measurements of $W$ and $W_i$ described in the text, along with the year of measurement and particle type.  Most data were taken with $\gamma$ and conversion $e^-$ sources, for which the relevant decay energy is listed. Ref.~\cite{Seguinot} used an electron beam with total energy listed below, and an energy per $e^-$ of 1--40~keV. Refs.~\cite{NEST_2011} and \cite{Akimov:2014cha} provide averages of subsets of previous measurements. }
\begin{ruledtabular}

  \begin{tabular}{ p{1.6 cm}  p{1.6cm}  >{\centering}p{3cm}  p{0.8cm} p{1.0cm} }
    $W$ (eV) & $W_i$ (eV) & Particle type &Year & Ref.\\  \hline
     -& 15.6$\pm$0.3 & $e^-$\,(976~keV) & 1975 &\cite{Takahashi}\\  
     -& 13.6$\pm$0.2 & $\gamma$\,(662~keV)& 1979 & \cite{Obodovski} \\  
    14.7$\pm$1.5&-& $e^-$\,(976~keV) &1990 &\cite{DOKE1990617,DOKE1993113} \\
    -& 9.76$\pm$0.70 & $e^-$\,(0.02-3~GeV) &1992 & \cite{Seguinot}  \\  
    13.8$\pm$0.9&-& $e^-$\,(976~keV) & 2002 & \cite{Doke}\\
    13.46$\pm$0.29&-&$\gamma$\,(122~keV) & 2007 & \cite{Shutt:2006ed} \\
    13.7$\pm$0.2&-&$\gamma$\,(122~keV, 136~keV) & 2009 & \cite{Dahl}\\
    14.0&-&$\gamma$\,(164~keV) & 2010 & \cite{2011APh....34..679A}\\
    13.7$\pm$0.4&-& Compilation & 2011 & \cite{NEST_2011}\\   
    -&16.5$\pm$0.8&$\gamma$\,(122~keV) & 2011 & \cite{Hornet}\\  
    -&14.30$\pm$0.14& Compilation & 2014 & \cite{Akimov:2014cha}\\
    
  \end{tabular}
  \end{ruledtabular}
   \label{prev-W}
\end{table}

The above summary indicates that there are substantial variations in previous measurements of the absolute calibration of charge and light yields of $\gamma$ and $\beta$ events in LXe. The spread in the values of $W$, either directly measured or derived from $W_i$, can be as large as 60\%.  Such differences could arise in part from detector effects such as variations in density, temperature, or xenon purity; differences in the energy deposition process for different particle types or energies; or unknown calibration systematics, as described in Sec.~\ref{sec:nest_compare}. While the relative calibration of charge and light yields versus field is straightforward for many detectors, the absolute measurement of these yields typically requires an accurate calibration of $n_q$, which can be performed in a single-phase detector through the use of a calibrated charge-sensitive amplifier. Since many modern large-scale LXe TPCs are dual-phase in order to amplify the charge signal prior to collection, absolute measurements of the charge yield are difficult due to the possibility that not all the charge is extracted into the gas phase, even at extremely high applied fields~\cite{PhysRevD.99.103024,Edwards_2018}.

Here we perform an absolute calibrated measurement of the total yield with a precision of 4.5\% using a variety of calibration sources with $\gamma$ energies between 1.1--2.6~MeV.  These measurements take advantage of a single-phase, large detector with good purity and a well-understood position and energy calibration, and the availability of a detailed Monte Carlo (MC) simulation of the detector energy response to calibration sources. 

\section{\label{setup}The EXO-200 experiment}
The EXO-200 detector was operated at the Waste Isolation Pilot Plant (WIPP) near Carlsbad, New Mexico from 2011--2018. The detector was filled with LXe enriched to $\sim$80\% in $^{136}$Xe, with a density at the operating temperature of 167~K of $3.0305 \pm 0.0077$~g/cm$^3$~\cite{EXO200}.  The LXe was housed in a cylindrical copper vessel, split into two TPCs by a common cathode, each with radius $\sim$18 cm and drift length $\sim$20~cm. Each end cap of the vessel consisted of two crossed wire grids and an array of large area avalanche photodiodes (APDs). More details of the detector are described in~\cite{EXOlatest} and~\cite{EXO200}. For each interaction, the charge was drifted parallel to the axis of the detector towards the nearest end cap under the action of a uniform electric field, and the scintillation light was collected and measured by the APD arrays. The drifting charge was measured by induced signals as it first drifted by a shielding, or ``V-wire" grid, and then was collected by a second wire grid, known as the ``U-wire" grid. 

In this work, we denote the $x$ and $y$ coordinates as those in the plane of the U- and V-wires, while the $z$ coordinate is defined to be along the drift axis of the detector, with the cathode at $z$=0, and the positive $z$ direction pointing from the anode of the second TPC to that of the first TPC. In each TPC, the V-wire grids were positioned at a separation $\Delta z = 6$~mm in front of the U-wire grids, while the APD arrays were positioned $\Delta z = 6$~mm behind the U-wires. The two wire grids were crossed at an angle of 60$^{\circ}$ and read out by charge sensitive preamplifiers. Each grid was segmented into 9~mm wide channels consisting of triplets of wires with 3~mm pitch. Each individual wire had a roughly square cross section with width of 127$\pm$40~$\mu$m, which enabled each wire grid to have an optical transparency of 95.8\%. 

The APDs were grouped into 74 readout channels (``gangs'') in total, each of which consisted of five to seven APDs. Each circular APD had a diameter between 19.6~mm and 21.1~mm and an active diameter of 16~mm. The APDs were hexagonally packed such that the sensitive area of the endplates on which they were mounted was 48\% of the total area. The interior surface of each endplate was covered by vacuum-deposited aluminum and MgF$_2$ to reflect VUV scintillation photons that did not strike the APD surfaces~\cite{EXO200}. A cylindrical PTFE reflector was positioned inside the electric field grading rings at a radius of 183~mm to improve light collection.

The V-wire and U-wire grids and the APD planes were separately voltage biased to ensure 100\% of charge was collected by the U-wires.  Simulations of the electric field and wire geometry indicate that all charge will drift through the V-wire plane and be collected on the U-wire plane for a ratio between the average electric field in the bulk of the TPC and collection region (between the U and V planes) $>1.3$.  To ensure full charge transparency, the grids were biased with a field ratio of 2 for all data considered here. Measurements performed during EXO-200 engineering runs, in which the field in the collection region was varied, confirm that this field ratio is sufficient to avoid loss of charge. 

At the event reconstruction stage, the signals on each channel are grouped into ``clusters'' based on their position and timing information.  The clustering algorithm is optimized to group signals arising from the same interaction into a single ``cluster'', from which the total energy and position for each energy deposit in the LXe can be determined.  Events that deposit their energy only in a single cluster are known as ``single-site'' (SS) events.  For simplicity, only SS events are considered for all analyses presented here, and ``multiple-site'' events where energy is deposited at multiple resolvable locations in the detector are not considered.  The SS events in the calibration data consist of either photoelectric absorption or closely-spaced, unresolved Compton scatters in the photopeaks for each source.  At energies below the photopeaks, the SS events consist of single, isolated Compton scatters with any additional energy deposits occurring outside the active volume of the detector.

\section{\label{yield}Measurement of scintillation and ionization yield using $^{228}$T\lowercase{h},  $^{226}$R\lowercase{a} and $^{60}$C\lowercase{o} sources}
\subsection{\label{yield1}Measurement procedure and results}

\begin{figure}[!t]
    \includegraphics[width=0.45\textwidth]{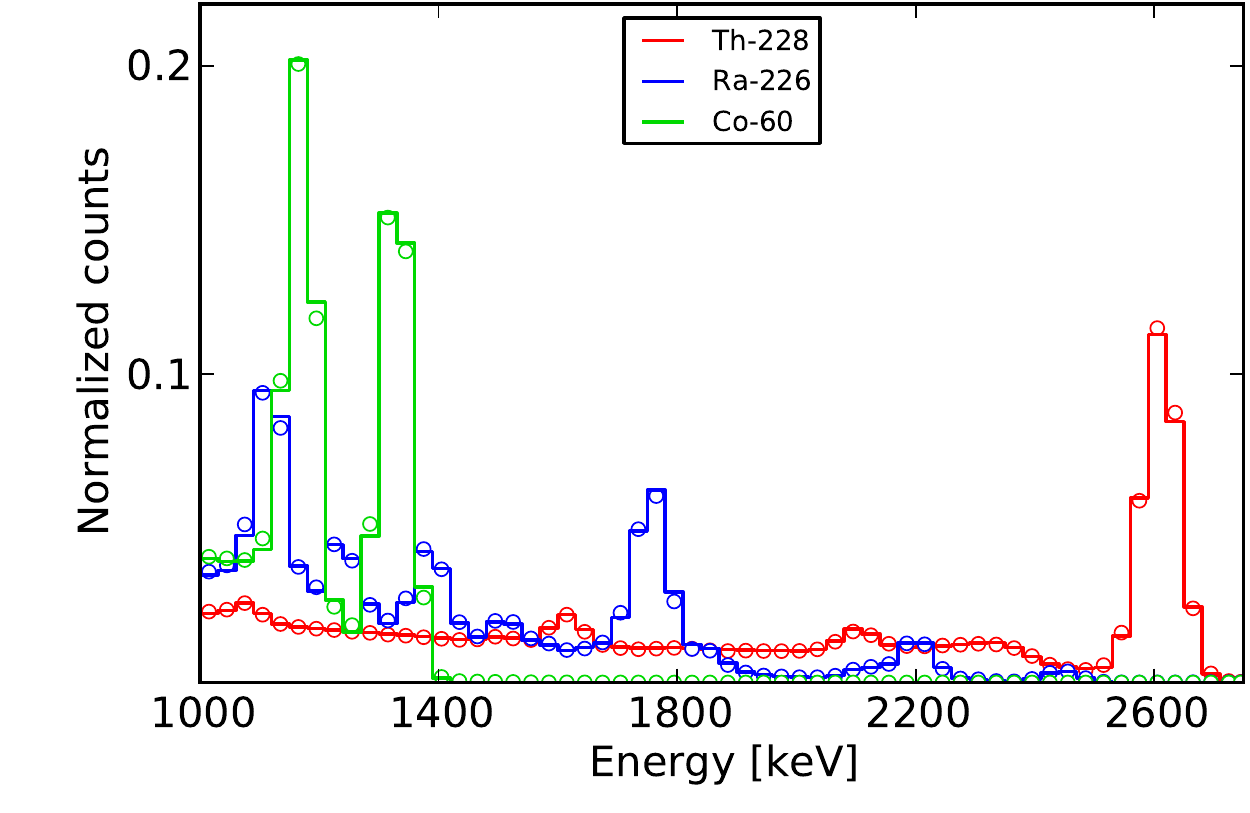}
    \centering
\caption{\label{fig:ii}Energy distributions in data (circles) and MC (lines) from $^{228}$Th, $^{226}$Ra, and $^{60}$Co calibration sources positioned near the cathode of the EXO-200 detector.}
\end{figure}

\subsubsection{Charge and light channel calibrations}
\label{sec:calibrations}
In order to measure the scintillation and ionization yield of LXe, we select data for analysis from $^{228}$Th, $^{226}$Ra, and $^{60}$Co source calibrations, with $\gamma$ signal energies of 2615~keV, 1764~keV, 1332~keV and 1173~keV (the latter are the energies of the two photopeaks of $^{60}$Co), respectively, as shown in Fig.~\ref{fig:ii}. For each calibration source, data were taken during two dedicated week-long calibrations: in February 2016 (near the end of EXO-200 ``Phase I'' operations~\cite{Albert:2014awa}) under electric fields 39~V/cm, 75~V/cm, 186~V/cm and 375~V/cm; and in October, 2018 (near the end of EXO-200 ``Phase II'' operations~\cite{EXOlatest}) under electric fields 50~V/cm, 100~V/cm, 200~V/cm, 400~V/cm and 567~V/cm.  Data were processed using the standard EXO-200 algorithms for event reconstruction and clustering described above.  In all the measurements, only events within the standard EXO-200 fiducial volume defined by the intersection of a hexagonal region with apothem $a<162$~mm and a cylindrical region of radius $r<173$~mm are considered. Along the drift axis, only events with $10$~mm~$<|z|<182$~mm are used~\cite{EXOlatest}. 

Since EXO-200 is a single-phase TPC, electrons are directly collected on the U-wires with high collection efficiency.  This allows an accurate determination of the total number of electrons produced in the LXe from an absolute calibration of the response of the charge readout electronics. The recorded waveforms from the charge and light channels in units of ADC counts are converted into electron and photon counts using the following calibrations.

\begin{figure}[t]
\centering
    \includegraphics[width=\linewidth]{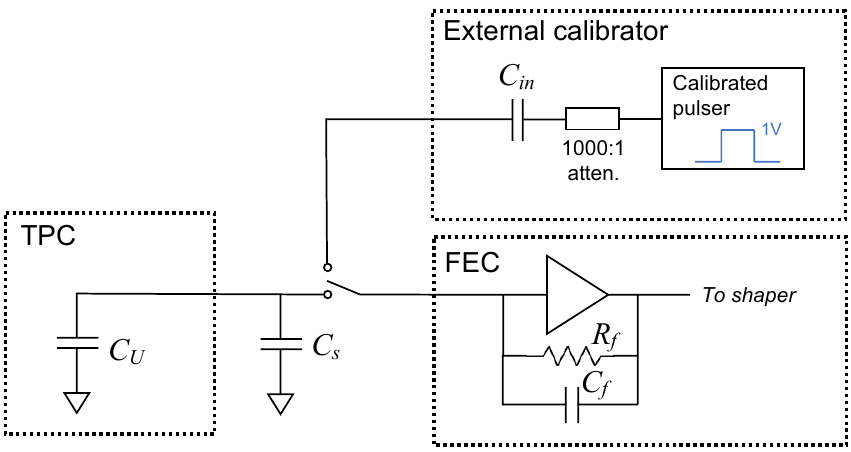}
\caption{\label{preamp}Schematic diagram of the external calibrator used for the ``external charge-injection'' calibration of the U-wire preamplifiers.  The external calibrator is used to apply a calibrated amount of charge to the input of the pre-amplifier through $C_{in}$.  The U-wire capacitance, $C_U$, and stray capacitance from the wiring, $C_s$, are disconnected during the calibration.  The FEC contains the charge sensitive preamplifier with $C_f = 1$~pF, $R_f = 60$~M$\Omega$, and open-loop gain $\gtrsim10^5$.}
\end{figure}

In the ``external charge-injection'' calibration, a known number of electrons from a calibrated capacitor are injected into the front-end electronics for the wire grids or APDs. The pulse magnitude recorded by the data acquisition (DAQ) system in units of ADC counts can then be determined for a calibrated amount of charge. 

Due to the importance of an accurate absolute calibration of the preamplifiers for the results reported here, the external charge-injection calibration for the U-wires was performed twice throughout EXO-200 operations. The calibration requires a specialized circuit and the detachment of the TPC from the front-end card (FEC) and therefore difficult to carry out frequently. Absolute charge-injection calibrations were originally performed for all U-wire, V-wire, and APD channels at the start of Phase~I operations.  To confirm the original calibration of the U-wires, the external calibration was repeated---for the U-wire channels only---using a newly fabricated external calibrator in November 2018 at the end of Phase~II running.  Both calibrations were found to agree on the average U-wire gain within 4\%. The difference between the two calibrations could arise from time variation in the gains over the course of detector operation, or from small systematic errors present in the Phase~I calibration, which was not designed to reach the percent level absolute accuracy of the Phase~II calibration. The Phase II calibration provides a measurement of the preamplifier response taken within a week following the calibration runs used here to avoid any systematics from changes in gain or operating characteristics throughout EXO-200 operations.

Fig.~\ref{preamp} shows a schematic of the circuit used for the Phase~II external charge-injection calibration for the U-wires. The test pulse is a step function with amplitude between 0.6--1.4~V and rise time much smaller than the sampling rate of the DAQ.  The absolute amplitude accuracy for the pulse generator was measured to be $<0.5$\%. The voltage step is applied through a calibrated attenuator with measured attenuation of $V_{out}/V_{in} = (1.00 \pm 0.01) \times 10^{-3}$ to the calibration capacitor, $C_{in}$.  This silver mica capacitor was hand-selected from tests of multiple capacitors to closely match the specified nominal value of 20~pF using a precision capacitance bridge (with absolute accuracy of $0.05\%$).  Measurements were performed before and after removing the capacitor leads, and after installing in the external calibrator board to account for any stray capacitance related to the installation. The total effective capacitance of the capacitor and board was found to be $C_{in} = 20 \pm 0.2$~pF, with the effects of stray capacitance once installed measured to be $\lesssim 0.1$~pF.  
 
The voltage step injected through $C_{in}$ provides a calibrated amount of charge ($\sim [75-195]\times 10^3\ e^-$) at the input to the preamplifier. The calibration is performed by disconnecting the U-wire cables coming from the detector from the FECs containing the charge preamplifiers.  When connected, the total detector and wiring capacitance, $C_s + C_U$, is dominated by the stray capacitance of the cables connecting the U-wires to the FECs, $C_s = 60-80$~pF~\cite{Auger}.  The feedback capacitance, $C_f = 1$~pF and open-loop gain of all preamplifiers $\gtrsim 10^5$ leads to a $<0.1$\% change in the amplifier response when the TPC and wiring are disconnected.  

Combining all systematic errors in the external calibration hardware described above, the gain uncertainty common to all channels is $\lesssim 1.5$\%.  Averaged over all channels, the resulting calibration indicates that each ADC count from a U-wire signal corresponds to $\sim$340 electrons, with a relative variation of 9\% across different channels. However, there are additional possible sources of systematic error resulting from possible time variation of these gains during data runs; differences in pulse shape between the charge injection calibration and physics data; and loss of electrons prior to collection by the U-wires.

To account for possible time variation in the gains, an ``internal charge-injection" calibration is also performed several times per week during the data-taking period. This calibration uses charge injected into the preamplifier directly from the calibration hardware on the FEC. Unlike the external charge injection, the total capacitance of the calibrator is not precisely known, so gains measured from the internal charge injection are used only to perform a relative measurement over time, which is anchored by the absolute value measured from the external charge-injection calibration. The internal calibration performed at the same time as each source calibration run is used to account for any time variation of the U-wire gain for that run.  The overall gain fluctuations are measured to be $\lesssim$1\% over the entire period of EXO-200 operations and $\lesssim$0.1\% over the week-long data taking period in Phase~II considered here. 

The charge injection calibrations are performed with step function input (with negligible rise time), while real physics signals have a charge pulse rise time varying between 3--6~$\mu$s~\cite{EXO200}.  This difference in rise time can lead to changes in the reconstructed amplitude of the signal after the pulse shaping electronics (i.e. ``ballistic deficit'').  Simulations show that this effect generates a relative $0.7$\% systematic error on the reconstructed pulse height for a signal with typical rise time relative to the step function input. 

Electrons in LXe can capture on electronegative impurities as they drift, which attenuates the charge signal. To minimize this attenuation, the xenon is continuously circulated through purifiers~\cite{Auger}. The purity of the LXe is monitored several times each week by dedicated source-calibration data, from which the electron lifetime can be determined. A drift-time dependent correction to the reconstructed charge energy is implemented in the data analysis, following the same procedure as previous EXO-200 analyses~\cite{2nubetabeta}. To limit effects of finite purity, only data for which the electron lifetime is $>$2 ms are used in this analysis.  The maximum charge loss for the lowest lifetime data considered here is $<5$\% over the full drift length, prior to correction.  After correction, the resulting error on the reconstructed charge is estimated to be $<0.5$\%.

Combining all systematic errors from the internal and external calibrations, electron lifetime calibration, and pulse shape studies, the total systematic uncertainty on the U-wire gain for each channel is 1.8\%.

Using this absolute calibration of the charge channels, the APD channels can be calibrated under the assumption of perfect recombination efficiency.  For this calibration, it is assumed that every recombined electron-ion pair produces an exciton, which de-excites to emit an additional VUV photon.  Under this assumption, the total number of quanta (either electron-ion pairs or photons) produced at a given energy is independent of field~\cite{Shutt:2006ed,Dahl}.  The change in the light and charge signals versus field can then be used to calibrate the total response of the light channels, which results from a product of the APD quantum efficiency, the geometrical collection efficiency, the APD avalanche gain, and the amplifier gain.  The advantage of this method is that the total photon count can be determined based only on the previously calibrated change in the number of electrons, and without the need to independently measure each component of the photon detection efficiency.  With additional calibration of the APD readout electronics it is possible to also determine the overall photon detection efficiency, $\epsilon_p$, defined as the ratio between the number of photon induced ``photo-electrons'' (PE) produced in the APDs prior to the avalanche amplification, relative to the total number of photons initially produced in the LXe.  The efficiency for detecting photons is significantly smaller than for charge due to the imperfect quantum efficiency of the APDs and the overall loss of photons as they are absorbed by uninstrumented detector surfaces.  

The external charge-injection calibration was also performed for the preamplifiers for each of the APD readout channels at the beginning of EXO-200 Phase~I operations.  While this calibration is not directly required to obtain the charge yields and $W$-value from EXO-200 data, along with the APD avalanche gain calibration, it can be used to estimate the photon detection efficiency from the calibrated total light response. The measured response indicates that one ADC count corresponds to $\sim$900 electrons at the input to the preamplifier, depending on channel.  The relative APD preamplifier gain variation is 11\% across different channels and the time variation is $\sim$1.5\% over the entire period of EXO-200 operations (excluding differences due to the electronics upgrade between Phase~I and II), as measured from the internal charge-injection calibration.  

The avalanche gain of each APD can be calibrated accurately using an {\em in-situ} laser calibration, during which all channels are illuminated by light from a pulsed 405~nm laser beam that enters the TPC through diffusers positioned at each end of the TPC.  The diffusers are illuminated by two optical fibers carrying light from an external laser source.  
Since the dynamic range of the DAQ is not sufficiently high to measure the response for both the unity gain (i.e. bias below the avalanche threshold) and full gain APD biases simultaneously, a two-step calibration is used where the laser pulse length is varied.  A short laser pulse is measured at full gain and at an intermediate gain and compared to a longer pulse measured at unity gain and the same intermediate gain. Both pulse lengths are short with respect to the times relevant to the front-end electronics. The ratio of the response amplitudes then gives the avalanche gain, which was measured during weekly calibrations throughout EXO-200 running.  The operating APD voltage biases for the data used here result in a mean gain of 200, consistent with the earlier EXO-200 measurement from~\cite{Neilson}. 
Combined with the preamplifier calibration, this corresponds to a conversion between APD pulse height and PEs on the order of 4.5~PEs/ADC.

\begin{figure*}[t!]
\centering
  \includegraphics[width=1.02\linewidth]{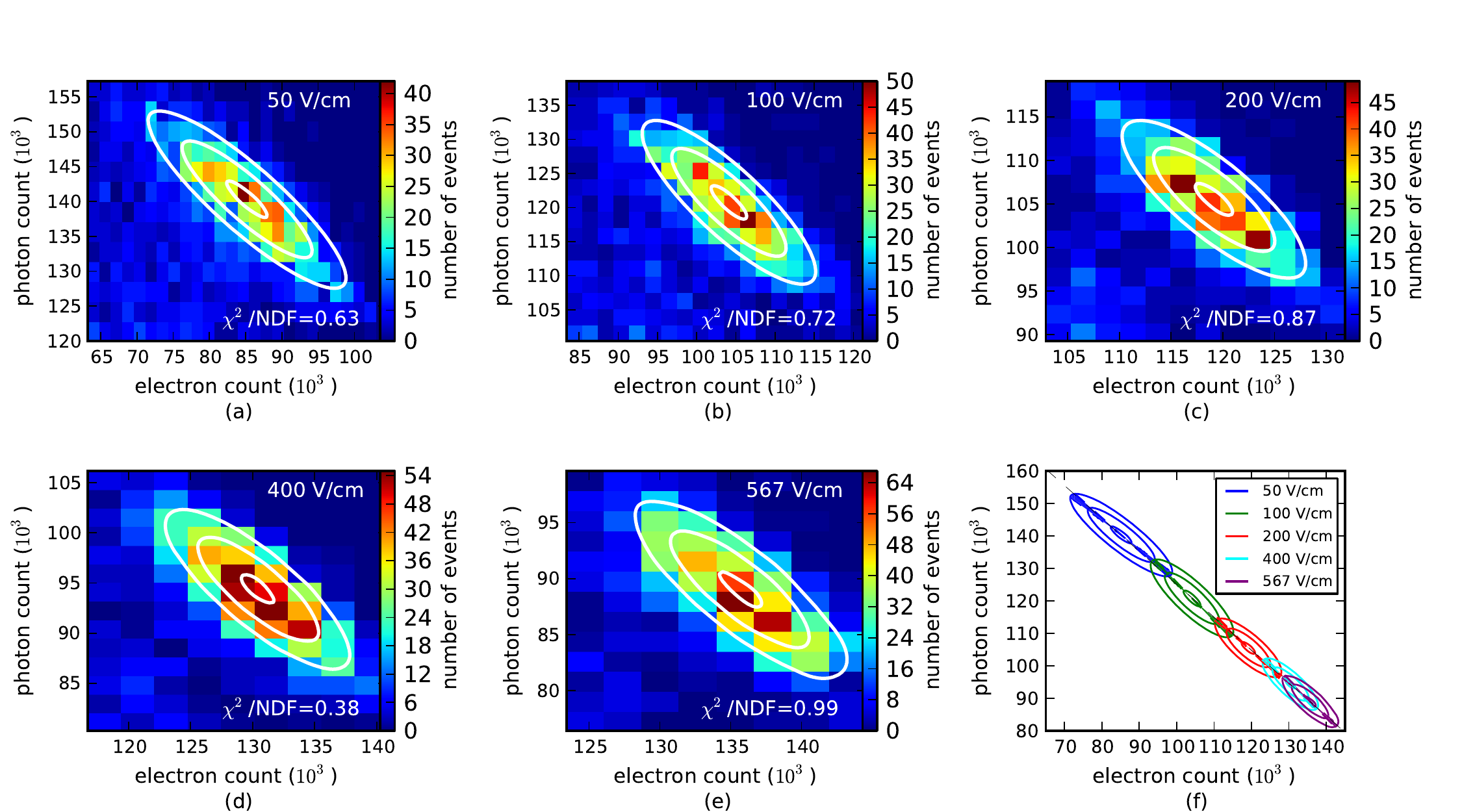}
  \caption{\label{gauss2d_Th}(a)-(e) MC-based fit to the anti-correlated number of electrons and photons at the 2615~keV $\gamma$ peak from the $^{228}$Th source data taken in October 2018 under various electric fields. Only bins with more than 10 events are included in the fit. The $\chi^2$/NDF fit statistic is indicated in each plot. The outermost contour contains 68\% of peak events on average for the best fit parameters. (f) Combination of the individual fits to the $^{228}$Th photopeak under various fields. The magnitude of the slope of each ellipses' major axis is equal to the tangent of the rotation angle. The shaded regions indicate the statistical uncertainty in the best fit value for the rotation angles.}
\end{figure*}
\subsubsection{Measurement of charge and light yields versus field}
\label{chargelightyield}

After obtaining the electron and PE counts in each event near the photopeak using the calibrations described in Sec.~\ref{sec:calibrations}, we perform a binned 2D Gaussian fit to the data: 
\begin{widetext}
\begin{equation}
\label{2dgauss}
f(x,y)=A e^{\frac{-1}{2(1-\rho^2)}\left[\frac{(x-n_q)^2}{\sigma_q^2}+\frac{(y-n_{PE})^2}{\sigma_{PE}^2}-\frac{2\rho(x-n_q)(y-n_{PE})}{\sigma_{PE}\sigma_q}\right]}
\end{equation}
\end{widetext}
where the $x$ and $y$ coordinates refer to the bin centers for the charge and light axes, respectively, $A$ is an overall amplitude, $n_q$ and $n_{PE}$ are the mean number of electrons and light-induced PEs at the photopeak, respectively, $\rho$ is the correlation between the electron and PE counts, and $\sigma_q$ ($\sigma_{PE}$) is the standard deviation of the electron (PE) counts, which includes both the detector noise and electron-ion recombination fluctuations. To account for the detailed energy spectrum of the calibration events, a simulation based on the EXO-200 detector MC~\cite{EXO200} is used to produce the expected event energy distribution for each source. This spectrum is smeared by a 2D Gaussian function for the resolution as in Eq.~\ref{2dgauss} to give the overall fitting function for each source.  The best-fit calibration and resolution parameters are then determined through a $\chi^2$ fit to the data. This MC-based fit accounts for events in the Compton shoulders near each photopeak, to minimize any effect of background events on the measurement. Using the best-fit values of electron and PE counts for each of the photopeaks in the calibration sources, the overall photon detection efficiency can be determined by requiring $n_q+(n_{PE}/\epsilon_p)$ to be a constant under different electric fields for each photopeak.

\begin{figure*}[t!]
\raggedleft
  \includegraphics[width=1.02\linewidth]{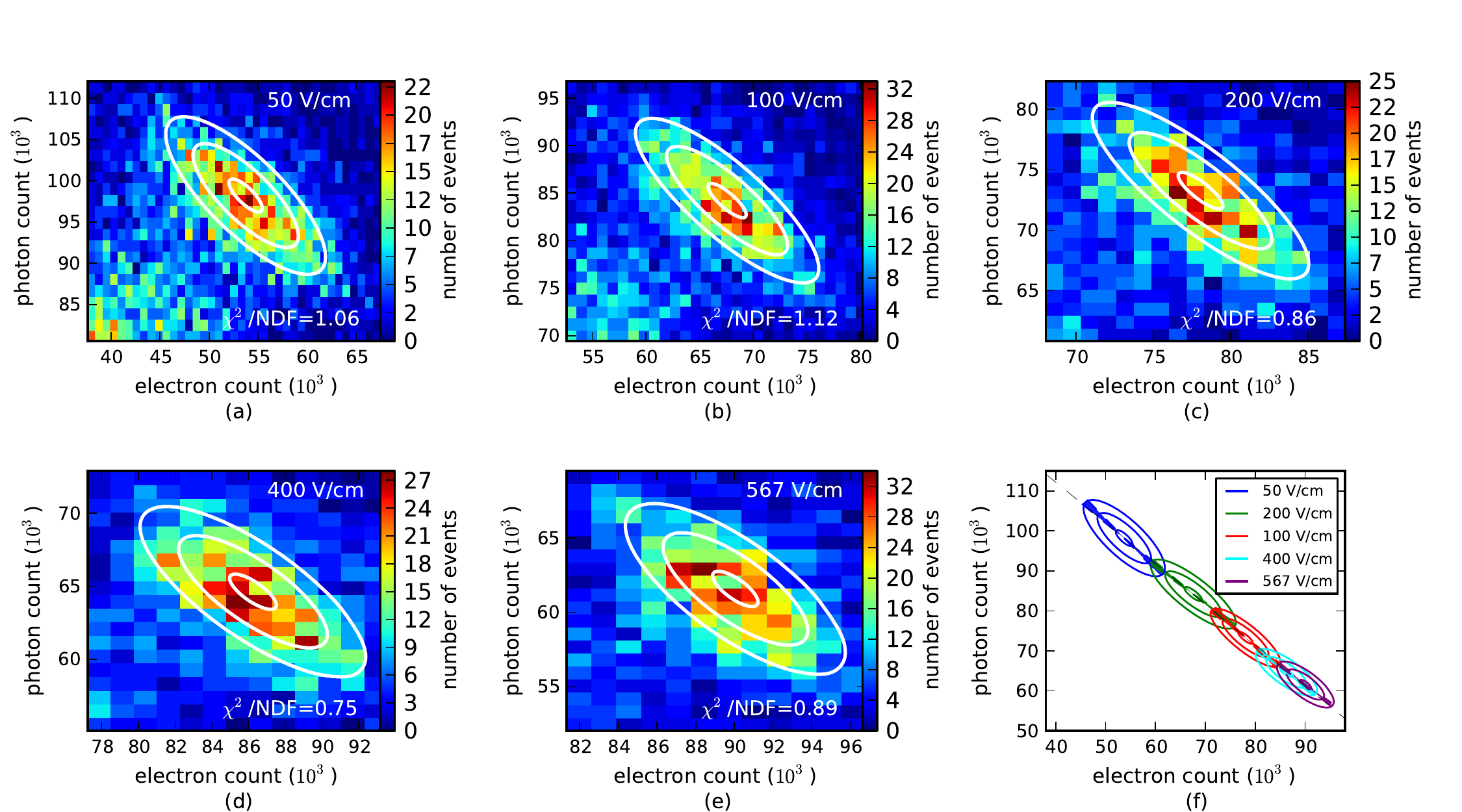}
 \caption{\label{gauss2d_Ra} Data and best fit to the 1764~keV $\gamma$ peak from the $^{226}$Ra source, following the same procedure as Fig.~\ref{gauss2d_Th}.}
  \end{figure*}

\begin{figure*}[t!]

  \includegraphics[width=1.02\linewidth]{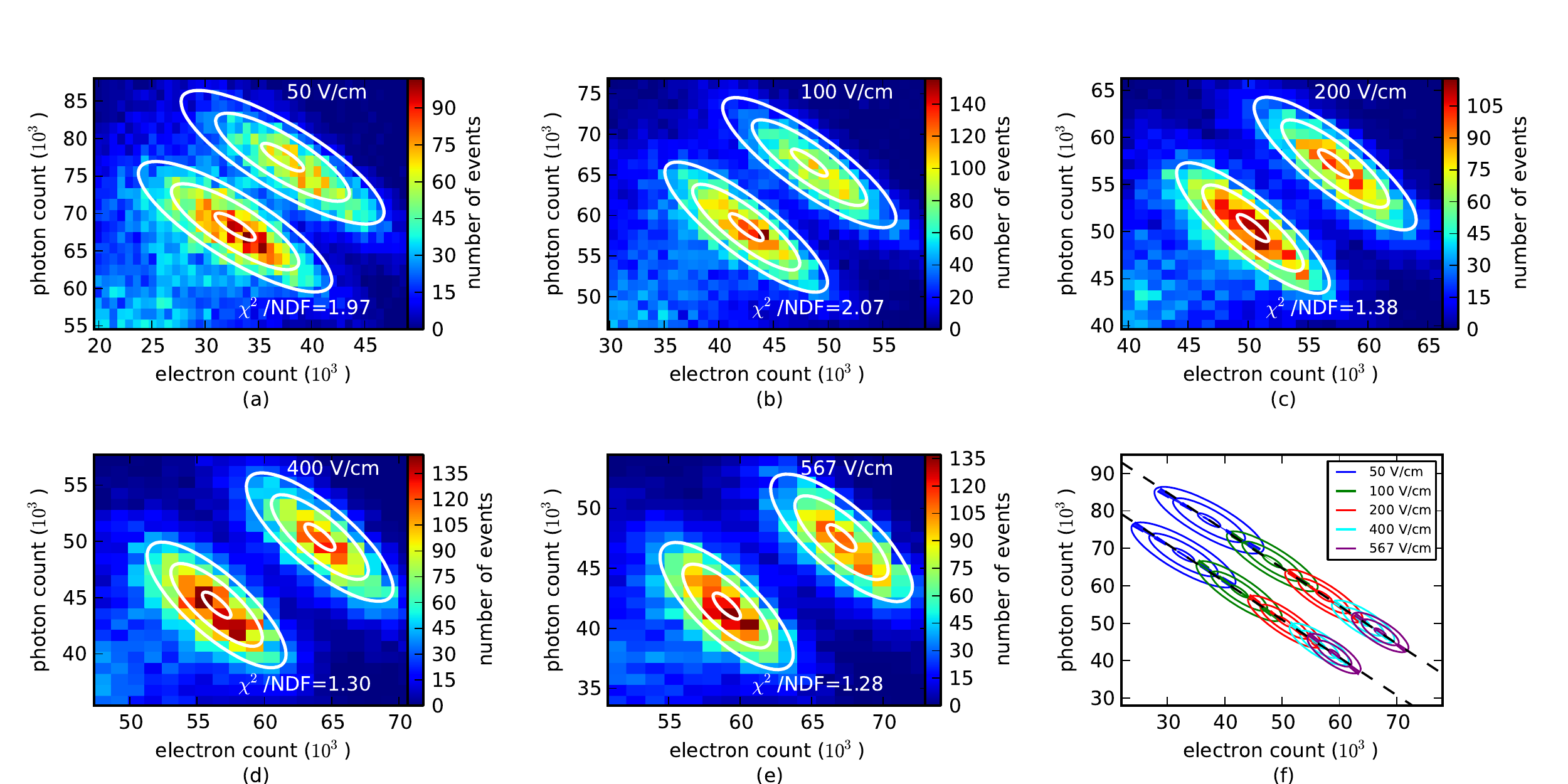}

\caption{\label{gauss2d_Co}Data and best fit to 1332~keV and 1173~keV $\gamma$ peaks from the $^{60}$Co source, following the same procedure as Fig.~\ref{gauss2d_Th}.}

\end{figure*}

The resulting photon detection efficiency estimated from the least squares fit to all calibration data from sources positioned near the cathode is $\epsilon_p$ = ($8.1 \pm 0.5$)\%, where the uncertainty is dominated by systematic variations between different calibration sources. Using this efficiency, which as described above relies on the assumption that each recombining electron-ion pair produces a photon, the best-fit values of $n_{PE}$ can be scaled to photon counts, $n_p$. 

\Cref{gauss2d_Th,gauss2d_Ra,gauss2d_Co} show the MC-based 2D Gaussian fit to electron-photon count spectra for the $^{228}$Th (2615~keV), $^{226}$Ra (1764~keV) and $^{60}$Co (1332~keV and 1173~keV) sources, respectively. Overall the use of the MC-based fit minimizes systematic errors due to the presence of backgrounds from Compton scattering near the photopeaks.  However, the difference between the results of the MC-based fit and a simple 2D Gaussian fit to each peak including a constant background is small ($\lesssim$5\%), indicating that background model systematics are not large.

The coefficient $\rho$ can be converted to the rotation angle, $\theta$, defined as $\tan(2\theta)=-\frac{2\rho\sigma_p\sigma_q}{\sigma_p^2-\sigma_q^2}$, indicating the optimal weighting of the charge and light signal to form the ``rotated energy'' that minimizes the overall resolution, as described in Sec.~\ref{res1}.

\begin{figure}[t]
\centering 
    \includegraphics[width=0.45\textwidth]{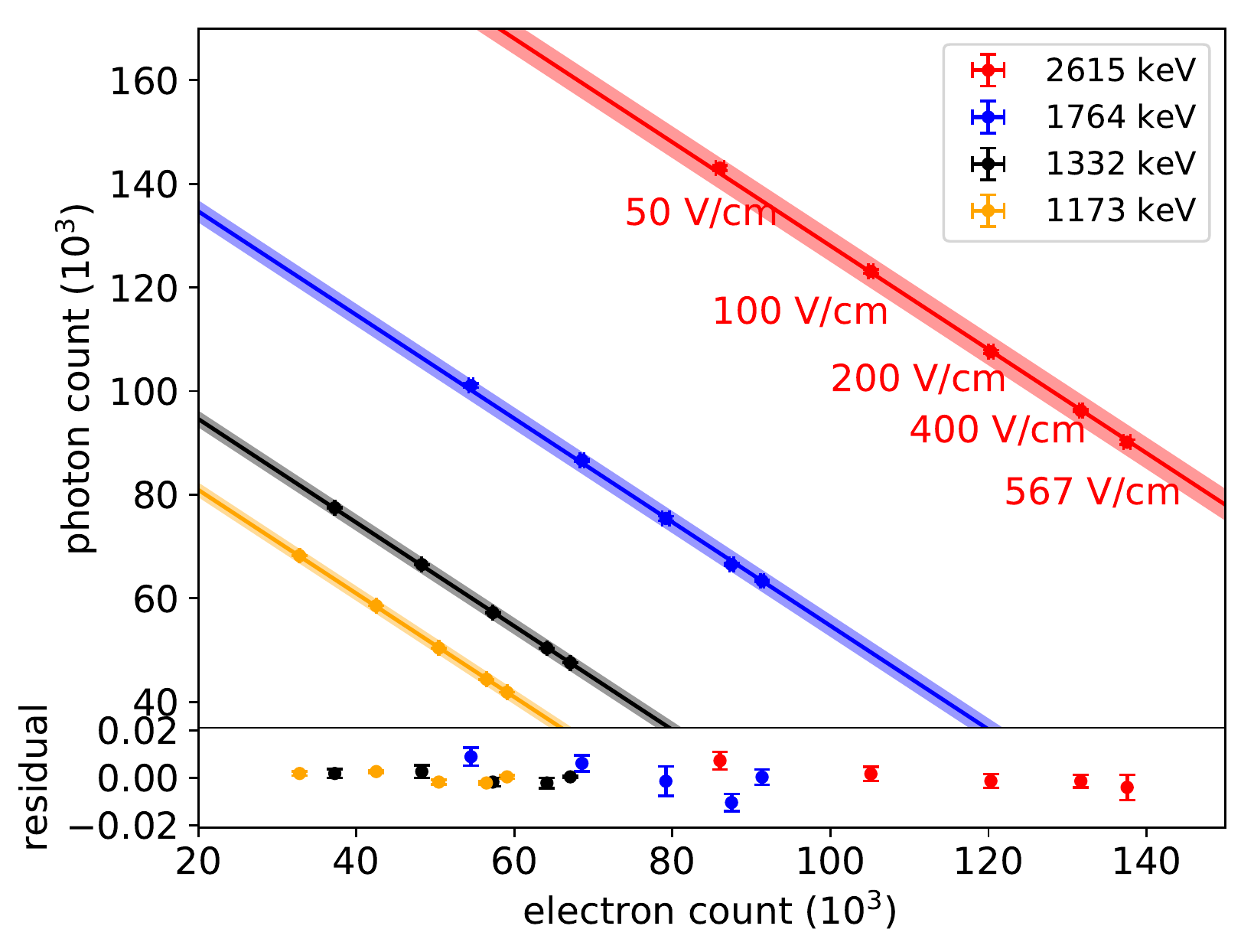}
  \caption{\label{fig:iii}Total number of photons vs. number of electrons at electric fields ranging from 50~V/cm to 567~V/cm, measured under the assumption that every recombining electron produces a photon.  The errors indicate statistical fluctuations from the fits shown in Figs.~\ref{gauss2d_Th}--\ref{gauss2d_Co}, while the shaded bands represent correlated uncertainties on the photon detection efficiency and APD/wire gain measurements. The photon detection efficiency is measured by requiring each line to have a slope of $-1$. Residuals between the data and the linear fit, i.e., ((data-fit)/fit) are indicated in the bottom panel along with the statistical errors.} 
\end{figure}

Figures~\ref{gauss2d_Th}f--\ref{gauss2d_Co}f show the total photon count versus electron count and the rotation angle $\theta$ under various drift fields measured using the three sources. The rotation angle decreases slightly as the electric field increases, due to the higher signal-to-noise in the charge channel compared to the light channel. In addition, as the drift field increases, the spread of the photon and electron distributions is reduced, leading to improved energy resolution. Fig.~\ref{fig:iii} shows the number of electrons and photons at the peak measured from various calibration sources under different electric fields. The estimated errors are dominated by the correlated uncertainties on the photon detection efficiency and APD/U-wire gain measurements.  To account for these uncertainties, an overall scale factor within the systematic error on the photon and electron count is indicated by the shaded bands.

From these data we measure the $W$-value defined in Sec.~\ref{prev_meas} as: 
\begin{equation*}
 W = 11.5\pm0.5 \mathrm{(syst.)} \pm 0.1 \mathrm{(stat.)}~\mathrm{eV}
\end{equation*}
The uncertainty on the $W$-value is dominated by systematic errors in the detector calibrations of the charge and light response.  The primary contribution to this systematic error is the 6\% uncertainty on the photon detection efficiency measured using the calibration sources at different electric fields.  The estimated 1.8\% absolute uncertainty on the charge response (described in Sec.~\ref{sec:calibrations}) provides a subdominant systematic error.

\begin{figure*}
\begin{minipage}[t]{0.48\linewidth}
\includegraphics[width=.9\linewidth]{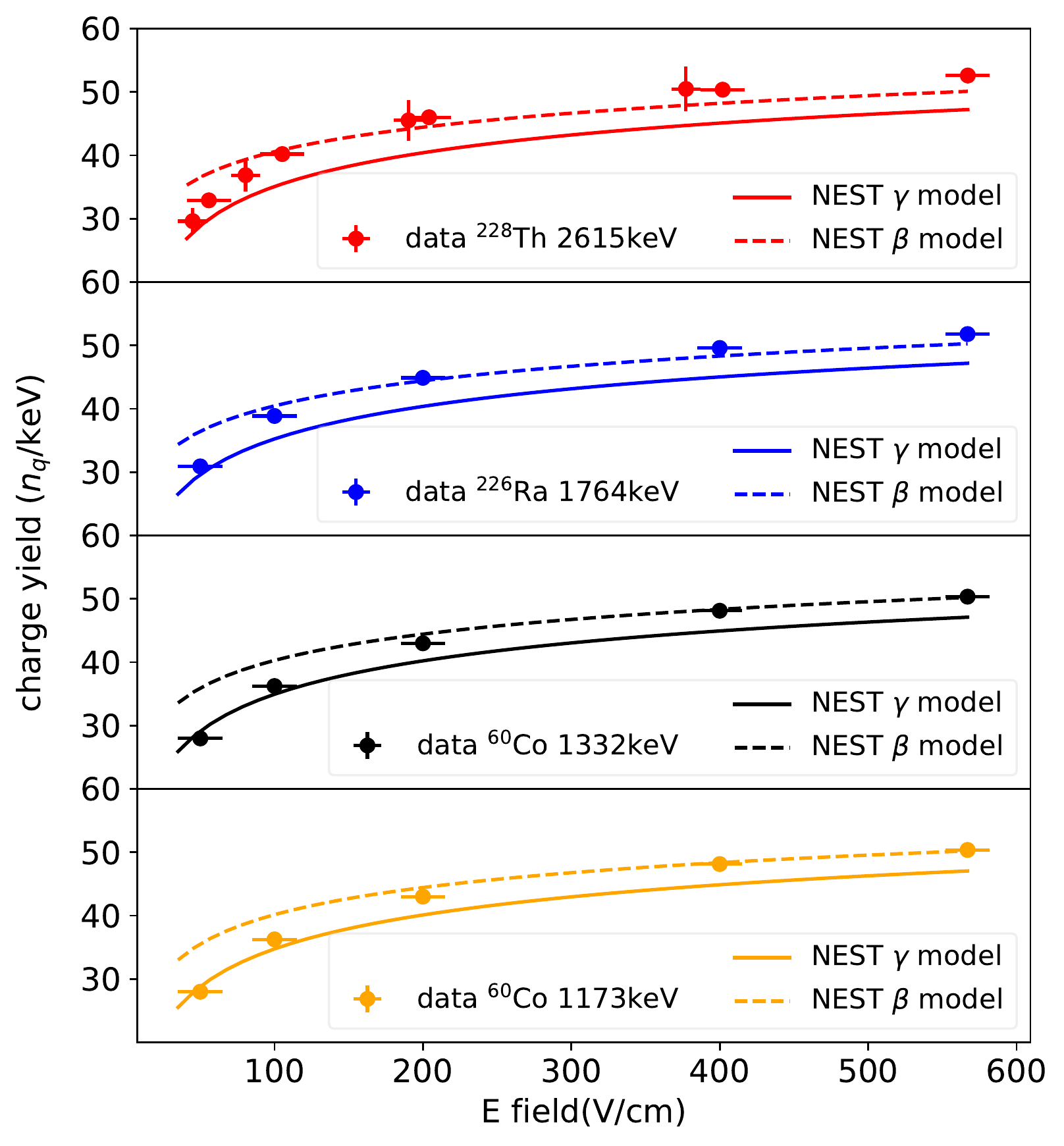}
\newline(a)
\end{minipage}\hfill%
\begin{minipage}[t]{0.48\linewidth}
\includegraphics[width=.9\linewidth]{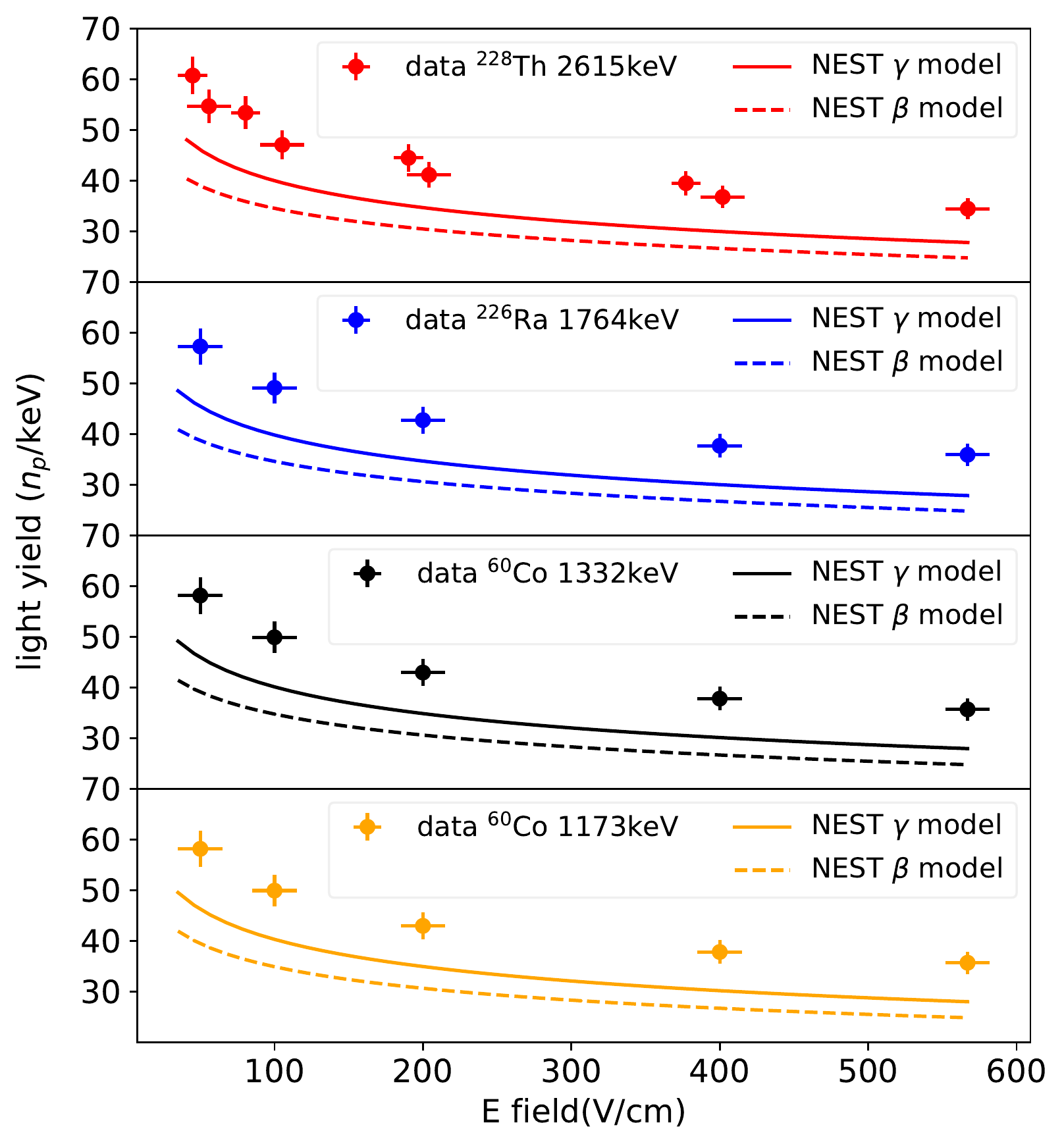}
\newline(b)
\end{minipage}%
 \caption{\label{NEST}Comparison of the measured charge yields (a) and light yields (b) at various electric fields with the NEST $\gamma$ and $\beta$ models. The light yields are measured under the assumption that every recombining electron produces a photon. For the $^{228}$Th source, measurements from both the Phase I and Phase II data sets are shown, while the other sources only have data available only from Phase II.}
\end{figure*}

\subsection{Comparison between experimental data and the NEST simulation}
\label{sec:nest_compare}

Fig.~\ref{NEST} shows the measured charge and light yield, defined as number of electrons, $n_q$, or photons, $n_p$, produced per keV of deposited energy for the three calibration sources.  The measured yields vary with electric field due to its effect on electron-ion recombination. The measured data are compared to the predictions from the NEST 2.0 simulation package~\cite{szydagis_m_2018_1314669}, with its $\gamma$ model (relevant for photoelectric absorption) and $\beta$ model (relevant for $\beta$ decays and Compton scattering) predictions shown as the solid and dotted lines, respectively. The NEST predictions are calculated for the density of the enriched Xe used in EXO-200 of 3.03~g/cm$^3$ as described in Sec.~\ref{setup}, and are simulated versus the electric field and energy for each of the photopeaks from the calibration sources.

For the $^{228}$Th data, calibrations from both the Phase~I and Phase~II data sets are shown, while the other sources only have data available for the Phase~II calibration.  The Phase~II data points have smaller errors due to their coincidence in time with the external charge calibration described previously.  Relative to this calibration, the Phase~I data have larger errors due to hardware modifications performed between the two data sets that introduce additional uncertainty on the absolute gain calibration in Phase~I.

The uncertainty on the electric field is determined by a 3D finite element simulation of the TPC geometry.  This simulation indicates that there is some position-dependence ($\lesssim 5$\%) in the magnitude of the electric field over the fiducial volume of the detector. In addition, evidence for charge buildup on the PTFE reflector surrounding the TPC is observed in the same datasets used here, at the lowest electric fields.  The estimated charge buildup is comparable to that seen in other LXe TPCs employing PTFE~\cite{Akerib:2017btb} and corresponds to a position dependent field distortion of 10--15~V/cm, independent of field.  This possible field distortion dominates the systematic error on the electric field in the detector, and is relatively more significant at lower fields.

\begin{figure*}[t]
\centering 
\includegraphics[width=1\textwidth]{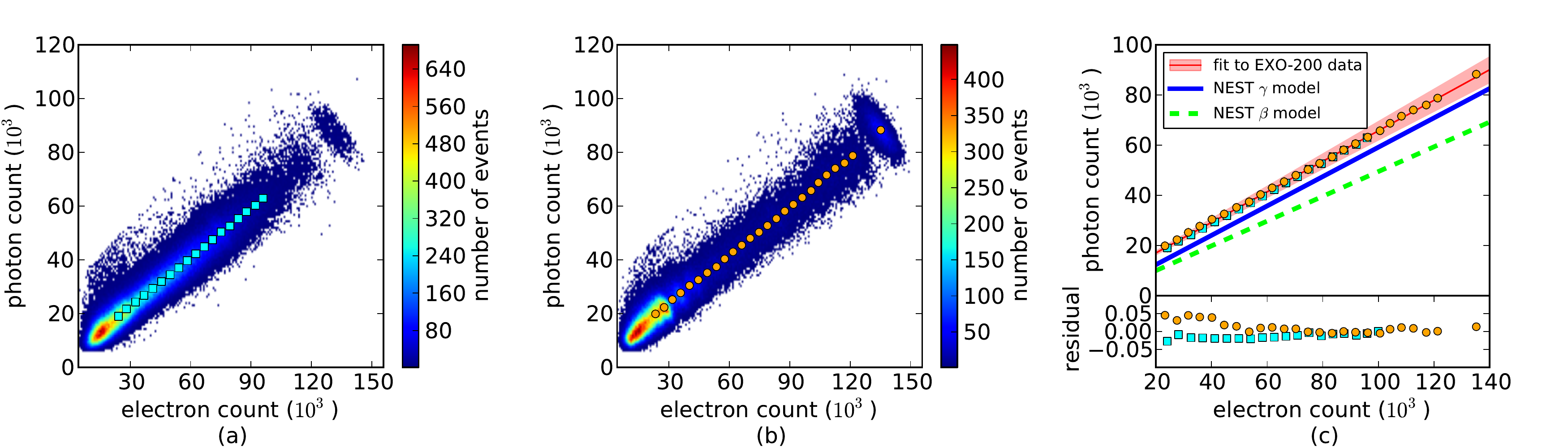}

 \caption{\label{LBandTh}(a)~Light versus charge response for events when no calibration source is present. Data within the continuous band are predominantly $2\nu\beta\beta$ events from $^{136}$Xe. The small peak at the end of the spectrum arises from residual backgrounds in the detector and is excluded from the fits. (b)~Charge and light response generated by $\gamma$-rays from the $^{228}$Th source. (c)~Average light versus charge response for $\gamma$ (orange circles) and $\beta\beta$ (blue squares) events at a range of energies ($>$500~keV) and an electric field of 567~V/cm. The difference between measurements and the linear fit to the combined dataset (indicated by the red line) in the energy range considered is within 5\%. The red bands show the 1$\sigma$ systematic errors on the response ratio, which are dominated by the uncertainty on $\epsilon_p$.  The solid blue and dashed green lines show a comparison with the NEST predictions. Note that the light responses in (a)-(c) are measured under the assumption that every recombining electron produces a photon.}
\end{figure*}

The EXO-200 detector also allows us to compare the charge and light yield for events with and without a calibration source present, as shown in Fig.~\ref{LBandTh}. In the absence of a calibration source, events are primarily generated by $^{136}$Xe $2\nu\beta\beta$ decays, which can be compared to the $\gamma$ events from the $^{228}$Th source at an operating field of 567~V/cm. The data without the source were acquired during Phase~II between July 2016 and October 2018, while the $^{228}$Th calibration dataset was taken within 2 days in October 2018. Only events in the fiducial volume within the energy range 500--2615~keV are used in the analysis. For both the $2\nu\beta\beta$ or $\gamma$ spectrum, data are first binned by rotated energy with non-overlapping bins of width equal to the 1$\sigma$ resolution at each rotated energy. In addition, a single bin encompassing the entire 2615~keV photopeak from the $^{228}$Th source is included.  This binning avoids bias in the selected energy of the events due to non-uniformity in the event distribution.  After binning in rotated energy, the charge and light yield at each point is determined from the median of the electron and photon counts in each bin and is plotted in Fig.~\ref{LBandTh}(c). A linear fit to the combined $2\nu\beta\beta$ or $\gamma$ dataset is indicated by the red line, and the residuals are shown in the bottom panel.

The ratio between the charge and light yield versus energy is found to be nearly constant over the energy range considered, with small deviations ($\lesssim 5\%$) occurring at low energies. In the energy range between 500~keV and 1500~keV, the charge-to-light ratio for events from the $^{228}$Th $\gamma$ events is $\sim$3\% smaller than those from $2\nu\beta\beta$ on average, and the difference grows larger for events with lower energy. No significant difference is found between the charge-to-light ratio for the $\beta\beta$- and $\gamma$-induced events in the energy range above 1500~keV. In particular, the charge-to-light ratio for single-cluster events from the 2615~keV photopeak of the $^{228}$Th source, which consists of both photoelectric absorption and closely spaced, unresolved Compton scatters, agrees with the corresponding average yield for $2\nu\beta\beta$ events within $\sim$1\%. The agreement between the charge and light yield for high-energy $\beta\beta$ and $\gamma$ induced events is consistent with the energy scale measured by fits to the detailed shape of the $2\nu\beta\beta$ spectrum in EXO-200. In these fits, the absolute $\beta\beta$ energy scale is found to be consistent with the calibrations using photopeaks from external $\gamma$ sources at the sub-percent level~\cite{EXO-200_final,EXOlatest}. 
In contrast to these data, NEST predicts a difference between the relative yields for its $\gamma$ model and $\beta$ model of $\sim$25\% in the charge-to-light ratio in this energy range, as indicated in Fig.~\ref{LBandTh}(c).

Relative to previous measurements, the data used here were taken at higher energies than most previous data listed in Table~\ref{table:1}.  While the data for the calibration sources considered here are consistent with a single value of $W$ between 1--2.5~MeV (i.e., no energy dependence is observed within this range), these data do not constrain energy dependence in this value below 1~MeV.
In addition, the EXO-200 APDs are sensitive to infrared red (IR) photons with wavelength $\lesssim$1000~nm, which could provide a difference in overall photon collection efficiency relative to experiments employing photo-multipliers (PMTs). While significant scintillation in the IR is observed in gas Xe~\cite{Belogurov,Bressi,Alexander_2016,Neumeier_2014}, IR emission in LXe has been measured to be substantially smaller~\cite{Bressi}. These previous measurements indicate that effects from additional collection of IR photons are expected to be small, although further work is needed to precisely measure the contribution from IR photons emitted in LXe~\cite{Alexander_2016}.

The NEST simulation of the charge and light yields has a small density dependence. At the EXO-200 enriched LXe density of 3.03~g/cm$^3$, NEST predicts a W-value of 13.3~eV. This value is 3\% smaller than the average value from Ref.~\cite{NEST_2011} of 13.7~eV, which corresponds to the value predicted by NEST at a density of 2.9~g/cm$^3$. Variations in the value of $W_i$ have been measured versus density in gas and liquid detectors~\cite{BOLOTNIKOV1997360}.  Additional density dependent effects are also parameterized in NEST for the parameter $\alpha$ and the charge and light yields based on global fits to previous measurements. However, including the higher mass density of the enriched LXe in the NEST simulation does not fully alleviate the differences between the data measured here and predictions.  In addition, for $\gamma$ and $\beta$ interactions, variations in the yields may be expected to vary with the electron density rather than mass density. Scaling the yields by the electron density instead would produce slightly larger tension between the EXO-200 measurements and the NEST predictions.

 In summary, our measurement of $W=11.5\pm0.5 \mathrm{(syst.)} \pm0.1 \mathrm{(stat.)\ eV}$ does not agree within errors with the NEST prediction. However, it does lie within the broad range of previous measurements summarized in Sec.~\ref{prev_meas}. The tension between our measurements and the NEST simulation cannot be fully relaxed by rescaling the NEST $W$-value down to 11.5~eV since the charge-to-light ratio predicted by NEST differs from our measurement. Differences as large as $\sim$8\% (6\%) in the charge  yield and $\sim$24\% (41\%) in the light yield are seen between the EXO-200 measurements presented here and NEST's $
 \gamma$ model ($\beta$ model) predictions, which is larger than the estimated systematic errors on these measurements. This is the first simultaneous measurement of absolute light and charge yields over the 1--2.5~MeV energy range, and can be used to improve modeling of this region in future iterations of the NEST software package.

\section{\label{rotationangle}Semi-empirical resolution model for EXO-200}
\subsection{\label{res1}Energy resolution model}
EXO-200 defines a ``rotated" energy scale
\begin{equation}
E \propto \cos(\theta)\langle E_q\rangle + \sin(\theta)\langle E_p\rangle
\end{equation}
where $\langle E_q\rangle$ is the estimated energy deposited as charge and $\langle E_p\rangle$ is the estimated energy deposited as light, for each event.  These energy estimates are determined from the amplitude
of signals observed in the summed charge and light channels:
\begin{equation}
  \langle E_{a}\rangle = \frac{A_{a}}{g_{a} \epsilon_{a}}W = n_a W
\end{equation}
where $A_a$ is the amplitude of the signal in ADC counts, $\epsilon_{a}$ is the average efficiency ({\em i.e.} fraction from 0-1) for measuring a given type of quanta, $g_{a}$ is the conversion factor between ADC counts and quanta, and $W$ is the average energy to create a single quantum (of either charge or light) for $a = (p,q)$. 

The total number of quanta can be estimated by:
\begin{equation}
  \label{eq:eng}
\langle n\rangle = \frac{\langle E\rangle}{W} \propto \cos(\theta)\langle n_q\rangle + \sin(\theta)\langle n_p\rangle
\end{equation}
We are generally interested in the relative energy resolution, which is given by $\sigma_E/E = \sigma_n/\langle n\rangle$, where $\sigma_n$ is the standard deviation of $\langle n\rangle$.  In terms of quanta, the relative variance can be expressed as:
\small
\begin{equation}\label{cov}
\frac{\sigma_n^2}{\langle n\rangle^2}= \frac{{\cos^2(\theta) \sigma_q^2 + \sin^2(\theta) \sigma_p^2 + 2\sin(\theta)\cos(\theta)\mathrm{Cov}_{q,p}+\sigma^2_{Xe}}}{\langle n\rangle^2}
\end{equation}
\normalsize
where $\sigma_{Xe}^2=f_{Xe}\langle n\rangle$ are the intrinsic fluctuations in the intial total number of quanta. The Fano factor $f_{Xe}$ is calculated to be $\sim$0.059 in LXe~\cite{Aprile}, but is typically sub-dominant to detector readout noise. 
Other sources of noise are also dominant in EXO-200, and even assuming a Fano factor as large as $f_{Xe} = 1$ would not lead to a significant change in the predicted resolution.  Thus, we exclude this factor in the following estimates.

The variance, $\sigma_q^2$, of our estimate of $n_q$ is:
\begin{equation}
\label{qnoise}
  \sigma_q^2 = \sigma_r^2 + n_q \frac{(1-\epsilon_q)}{\epsilon_q} + \frac{\sigma_{q,noise}^2}{\epsilon_q^2}
\end{equation}
where $\sigma_r^2$ is the variation in units of quanta resulting from recombination fluctuations. The second term represents the binomial fluctuations for non-unity charge collection efficiency, and $\sigma_{q,noise}^2$ is the electronics noise of the charge collection wires in units of electrons.  Since the average electron lifetime of the selected data is 3.2~ms, the fraction of electrons absorbed by impurities in the LXe is $\sim$3\% and therefore the charge collection efficiency $\epsilon_q = 97$\%. Due to this high collection efficiency, the second term in Eq.~\ref{qnoise} is negligible compared to the electronics noise and recombination fluctuations.

Similarly, we can write the variance of the estimator of the number of photons as:
\begin{equation}
\label{sigmap2}
  \sigma_p^2 = \sigma_r^2 + \frac{n_p}{\epsilon_p}[(F_N-1)+B^2]+ \frac{\sigma_{p,noise}^2}{\epsilon_p^2}+n_{p}^2\sigma^2_{NU}
\end{equation}
In addition to the recombination fluctuations, $\sigma_r^2$, and the electronics noise of the APD readout channels, $\sigma_{p,noise}^2$, there are three additional
noise terms caused by fluctuations related to the initial number of PE created in the APDs, $n_{PE} =
\epsilon_p n_p$. The term $B^2\cdot n_p \epsilon_p$ describes fluctuations in the number of PEs created by the photons reaching the APDs. The factor $B$ is determined from the binomial fluctuations in the number of detected photons, due to the imperfect collection efficiency, and also includes sub-Poissonian fluctuations arising from the creation of PEs by VUV photons in the Si APDs.  

For 5.9~keV X-rays interacting in Si, the PE creation process has been measured in detail~\cite{LOWE2007367, Neilson}. These measurements indicate that the average energy to create a PE is 3.72~eV~\cite{LOWE2007367} at LXe temperatures, with the fluctuations in $n_{PE}$ well described by a Fano factor of $\sim$0.1.  However, at the much lower energy corresponding to 178~nm VUV photons (7.0~eV per photon), the intrinsic PE creation process and the number of PEs ($\mu_{PE}$) created by a single photon in Si is less well-characterized.  A direct measurement of PE creation by 7.0~eV photons found a mean of $\mu_{PE} = 3.0$~PE/photon~\cite{PhysRev.134.A205} with an 8\% uncertainty, which implies a lower creation energy per PE than for keV-scale X-rays.  For $\sim$5~eV photons, measurements have found values between 1.3--2.0 PE/photon~\cite{VAVILOV1959223,wolf98,doi:10.1063/1.4819447,doi:10.1063/1.110489}, with the higher end of this range being consistent with the measurements in Ref.~\cite{PhysRev.134.A205}. Simulations of PE creation by VUV photons are consistent with a mean of $\mu_{PE}=$2--3 PE/photon at 7~eV~\cite{2010NJPh...12g3037M,BRIGIDA2004322}, with variance described by a Fano factor, $f_{Si}(\mathrm{7~eV}) \approx 0.2$~\cite{BRIGIDA2004322}.  

To determine the overall value of $B$, which includes fluctuations in both the number of photons detected as well as fluctuations in the PE creation process, a two-step simulation is employed.  First, a random number of collected photons is determined from a binomial distribution with $n_p$ trials and probability $\epsilon_p/\mu_{PE}$.  For each collected photon, a discrete number of PEs is then generated from a distribution with mean $\mu_{PE}$ and variance of $f_{Si}\mu_{PE}$.  The total number of PEs, $n_{PE}$, is then determined for each simulated event along with the variance over all trials.  Under all electric fields, and for $\mu_{PE}$ varying from 2--3, the value of $B$ is $1.8 \pm 0.2$, where the error accounts for the uncertainty in $\mu_{PE}$.

Following Ref.~\cite{Neilson}, there are two additional  
variance terms relative to the quanta $n_{PE}$ at the input to the APD:
$\sigma_{PE}^2 = (F_N - 1) n_{PE}+n_{PE}^2\sigma^2_{NU}$. The first term is related to fluctuations of the APD avalanche gain that are parameterized by the excess noise factor $F_N$~\cite{Neilson,MOSZYNSKI2002504}. The second term accounts for non-uniformity or position dependence of the detector response.  This non-uniformity can arise both from differences in gains between APD gangs and differences in gains between the APDs within a single gang. The overall variation in gain between different gangs and over time can be calibrated and removed using the source calibration data~\cite{Davis:2016reu}.  While this overall gain variation can be $\gtrsim$10\%, $\sigma_{NU}$, which represents the variation in the total light response is measured to be only $\sim1$\% for events from sources near the cathode, where the uniform distribution of photons averages over these gain variations.  For the data considered here, the contribution to $\sigma_{NU}$ due to gain variations between gangs is sub-dominant compared to other terms in the resolution model, even prior to applying the gang-dependent correction. 

In addition to the gain non-uniformity among various APD channels, there may also exist gain non-uniformity within an APD channel. The APDs within each gang were selected to have matched gains based on testing prior to installation~\cite{Neilson}, but small residual differences remain. Moreover, slight time variations in the gains of an individual APD are possible due to changes in temperature and other systematic effects. While the overall gain is calibrated for each gang as a function of time, gain non-uniformity within the gang cannot be calibrated and could lead to additional variation in the light response. This non-uniformity is studied in detail in Appendix \ref{APDnon}, and is found to not significantly impact the resolution for events in the fiducial volume of the detector.

Source calibration data can be used to determine the relative number of photons collected for interactions at different locations in the detector. This position-dependent response is used to calculate the ``lightmap,'' which describes the summed response of all APDs as a function of event position and time~\cite{EXO200,Davis:2016reu}. Since the lightmap is constructed empirically from calibration data from the $^{228}$Th source, detector regions far from the calibration source can have limited statistics, leading to an uncertainty on the detector response.  Such statistical or systematic errors in the lightmap can lead to position-dependent errors in the energy estimate.  While it is difficult to simulate the light response of the EXO-200 detector at the percent level accuracy needed to verify the empirical lightmap, as will be shown in Sec.~\ref{res2}, the resolution model can describe the experimental data without including additional sources of position dependent error.  The agreement of the measured resolution and model indicate that systematic errors in the lightmap are subdominant compared to other sources contributing to the energy resolution.

Finally, the covariance between the light and charge signals is represented by $\mathrm{Cov}_{q,p}$.  Assuming perfect recombination efficiency, i.e. that every recombined electron results in the emission of a VUV photon, the recombination fluctuations in the charge and light signals will be identical, as indicated in Eq.~\ref{qnoise} and \ref{sigmap2}, and the covariance between electron and photon counts will be $\mathrm{Cov}_{q,p}=-\sigma_r^2$.  As described below, consistency of this model with the measured resolutions can provide constraints on the assumption of perfect recombination efficiency.

\begin{table}[t]
\caption{ Measured values for quantities independent of the drift field in the resolution model and their estimated errors}
\begin{ruledtabular}

  \begin{tabular}{ p{3.5 cm}  p{1.2cm}  p {1.4cm}  p{1.4cm} }
    Quantity & Value & Syst. err. [\%] & Stat. err. [\%] \\  \hline
    $\epsilon_p$ & 0.081 & 6.2 & 0.6 \\  
    $F_N$ & 2.15 & 11.9& 1.4 \\  
    $\sigma_{q,noise}$($e^{-}$)& 770 & 4.5 & 1.5 \\  
    $\sigma_{p,noise}$(PE) [Phase~I/II]& 446/148& 3.5 & 0.5 \\  
    $\sigma_{NU}$ & 0.012& 5.0 & 5.3 \\  
  \end{tabular}
  \end{ruledtabular}
   \label{table:1}
\end{table}

\begin{table}[t]
 \caption{ Quantities depending on the drift field measured using 2615~keV $\gamma$s from the $^{228}$Th source. Systematic and statistical errors are included.}
\begin{ruledtabular}
  \begin{tabular}{  p{1cm}  p{2cm}  p{1.9cm}  p{1.9cm}  p{1.9cm} }
    E~field (V/cm)&$n_q (\times 10^3)$ (syst.)~(stat.) &$n_p (\times 10^3)$ (syst.)~(stat.)& $\sigma_r (\times 10^3)$ (syst.)~(stat.)\\  \hline
    39 & $79 \pm 5.5 \pm 0.1$   & $161 \pm 9.7 \pm 0.1$     &$8.4\pm0.5\pm0.3$\\ 
    50 & $86\pm1.5\pm9.8$   & $143\pm8.9\pm6.6$         & $8.3\pm0.4\pm0.5$\\ 
    75 & $98\pm6.9\pm0.1$   & $141\pm8.5\pm0.1$     &$7.8\pm0.5\pm0.3$\\ 
  100  & $105\pm1.9\pm0.7$ & $123\pm7.6\pm0.9$  &$7.1\pm0.3\pm0.4$\\ 
  186 & $121\pm8.5\pm0.1$   & $118\pm7.1\pm0.1$     &$6.1\pm0.5\pm0.2$\\ 
  200  & $120\pm2.2\pm0.5$ & $107\pm6.6 \pm0.5$ &$5.4\pm0.3\pm0.4$\\  
 375 & $134\pm9.4\pm0.1$   & $105\pm6.3\pm0.1$     &$5.0\pm0.5\pm0.2$\\ 
400 & $132\pm2.4\pm0.4$  & $96\pm5.9\pm0.5$      &$4.9\pm0.2\pm0.3$\\  
567  & $138\pm2.5\pm0.4$ & $90\pm5.6\pm0.5$     & $4.7\pm0.2\pm0.3$ \\ 
615 & $141\pm9.9\pm0.1$   & $97\pm5.8\pm0.1$     &$3.6\pm0.8\pm0.3$\\ 
 
  \end{tabular}
  \end{ruledtabular}
   
  \label{table:2}
\end{table}

\subsection{\label{res2}Optimal energy resolution predicted by the model}
The quantities $n_q, n_p, \epsilon_p, \sigma_{q,noise},\sigma_{p,noise}, \sigma_r$ and $F_N$ in the resolution model are directly measured from experimental data. The measurements of $n_q, n_p$, and $\epsilon_p$ have been described in Sec.~\ref{yield1}. The total electronics noise for the APD and U-wire channels is measured by fitting the pre-pulse baselines recorded by the DAQ to the same signal model used to reconstruct data. This method ensures that the reconstructed noise is filtered and processed in the same way as the detector signals.  The resulting noise measurements, after accounting for the average channel multiplicity (i.e. that multiple charge and light channels typically need to be summed to fully reconstruct all energy) are shown in Table~\ref{table:1}.  The charge channels have similar noise in Phase~I and Phase~II data, with a mean of $\sigma_{q,noise} = 770$~$e^-$ and $\sim 4.5$\% variation over different channels.  The light channels have significantly smaller noise in Phase~II relative to Phase~I due to an upgrade to the electronics between the two operating periods~\cite{EXOlatest}.  The measured noise summed over all APD channels is $\sim$450~PEs in Phase~I and $\sim$150~PEs in Phase~II, corresponding to a reduction by $\sim3\times$.

\begin{figure}[t]
\centering 
    \includegraphics[width=0.48\textwidth]{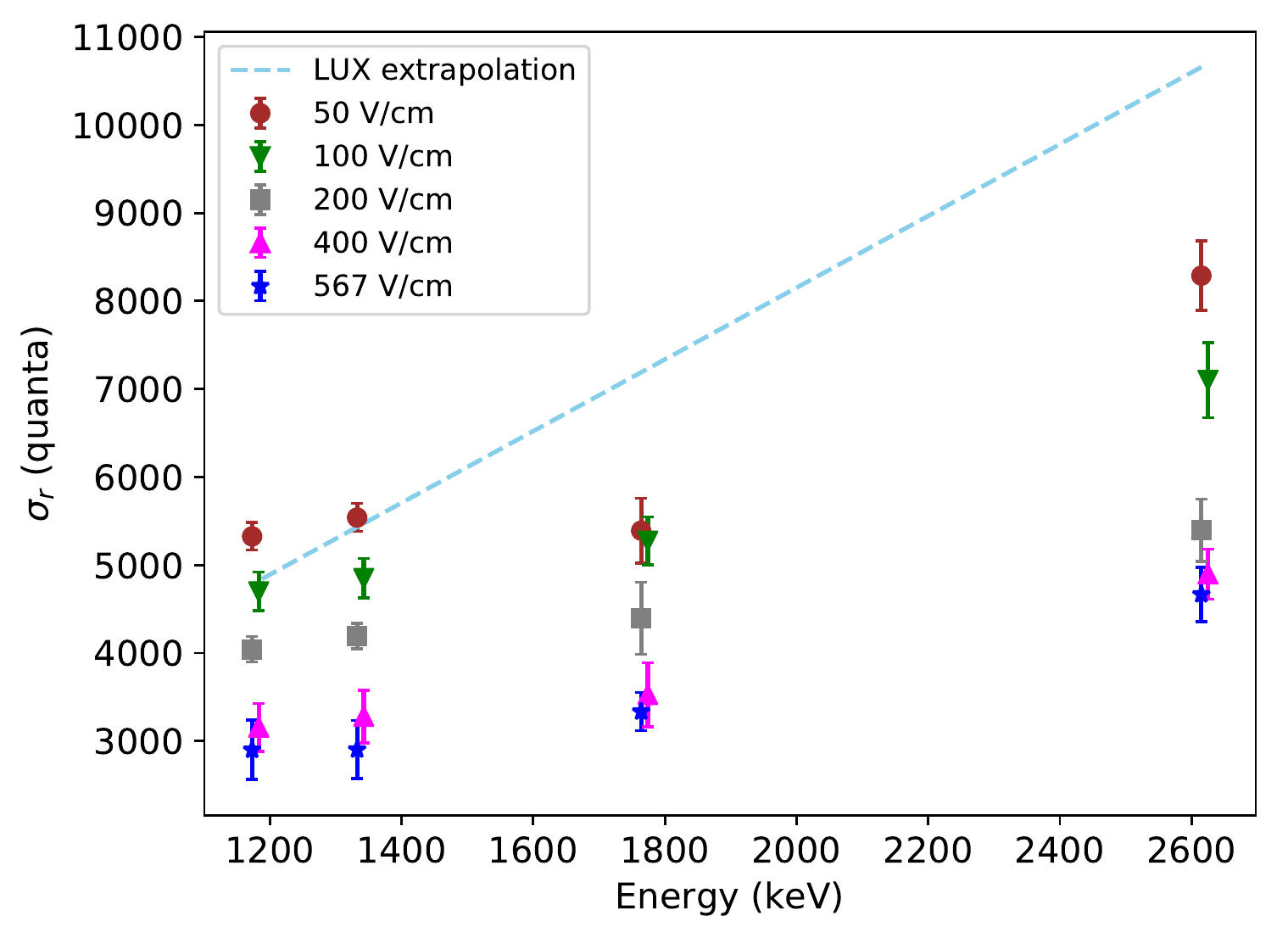}
 \caption{\label{sigmarvs}Measured recombination fluctuations versus incident $\gamma$-ray energy under different electric fields. All points are taken at the energies corresponding to the photopeaks of the calibration sources, but a small plotting offset is added in energy to improve visibility of their errors. The dashed line is a linear extrapolation from the LUX measurements performed at fields varying from 43 to 491~V/cm using $\beta$ decay signals~\cite{Akerib:2019jtm}.}
\end{figure}

The recombination fluctuations, $\sigma_r$, listed in Table~\ref{table:2} are measured by subtracting the detector noise from the total measured variance in the charge and light photopeaks~\cite{Dobi:2014wza}. 
Fig.~\ref{sigmarvs} shows the measured recombination fluctuations, $\sigma_r$, as a function of energy using data taken during Phase II. While these recombination fluctuations have been found to scale approximately linearly in energy at energies below 1~MeV~\cite{Akerib,Akerib:2015wdi,Akerib:2019jtm}, in the higher energy region measured here, the increase in $\sigma_r$ with energy is found to be smaller than would be expected from extrapolating the linear dependence observed at lower energy.  

The same laser calibration data used to determine the APD avalanche gain (described in Sec.~\ref{sec:calibrations}) can be used to measure the APD excess noise factor, $F_N$, due to fluctuations in the avalanche process.
The variance of the amplitude of repeated laser pulses (5000 pulses are taken at each gain setting in each calibration) can be written as $\sigma_{laser}^2 = G^2F_N n_{PE}+n_{PE}^2\sigma^2_{NU}+\sigma_{noise}^2$, where $G$ is the APD gain obtained in Sec.~\ref{yield1} and $n_{PE}$ is the average number of PEs per pulse.  Here $\sigma_{noise}$ includes all noise terms that do not depend on the avalanche gain, including electronics noise, fluctuations in the laser power between pulses, etc.  Since the laser light passing through the diffuser in the opposite TPC illuminates each APD channel in a gang approximately uniformly, the non-uniformity term $\sigma^2_{NU}$ is negligible. Therefore the excess noise after avalanche can be estimated using the measured variance in the unity gain calibration, $F_N = (\sigma_{laser}^2-\sigma_{p,noise}^2)/(G^2 n_{PE})$.

The values of the resolution model quantities that are independent of the electric field are listed in Table~\ref{table:1}. The measured electric field dependent quantities are listed in Table~\ref{table:2}.

\begin{figure}[t]
\centering 
    \includegraphics[width=0.48\textwidth]{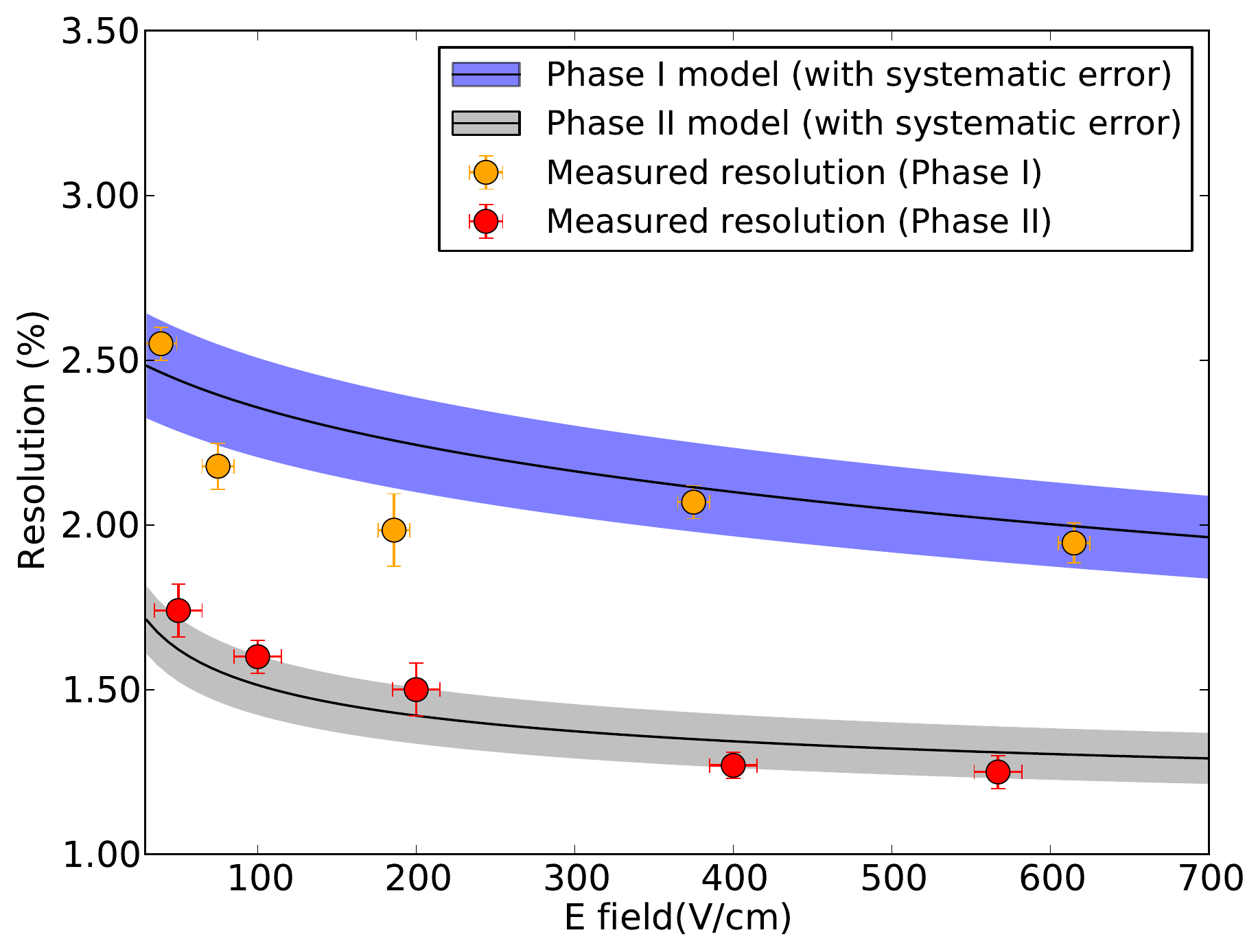}
  \caption{\label{ang}Comparison between measured rotated energy resolution at 2615~keV under different electric fields and predictions from the resolution model. Data taken during Phase~II has better resolution than Phase~I due to the reduced APD noise after the electronics upgrade.}
\end{figure}

 Using the measured quantities above, the best resolution predicted by the model can be obtained through minimizing Eq.~\ref{cov} with respect to the rotation angle $\theta$. The comparison between the measured resolution in Phase~I and Phase~II, and the predictions from the resolution model under various electric fields is shown in Fig.~\ref{ang}. The resolution values shown here are measured without implementing the de-noising algorithm described in~\cite{Davis:2016reu} to allow the noise to be directly estimated from the summed charge and APD waveforms.

\begin{figure}[t]
\centering 
    \includegraphics[width=0.48\textwidth]{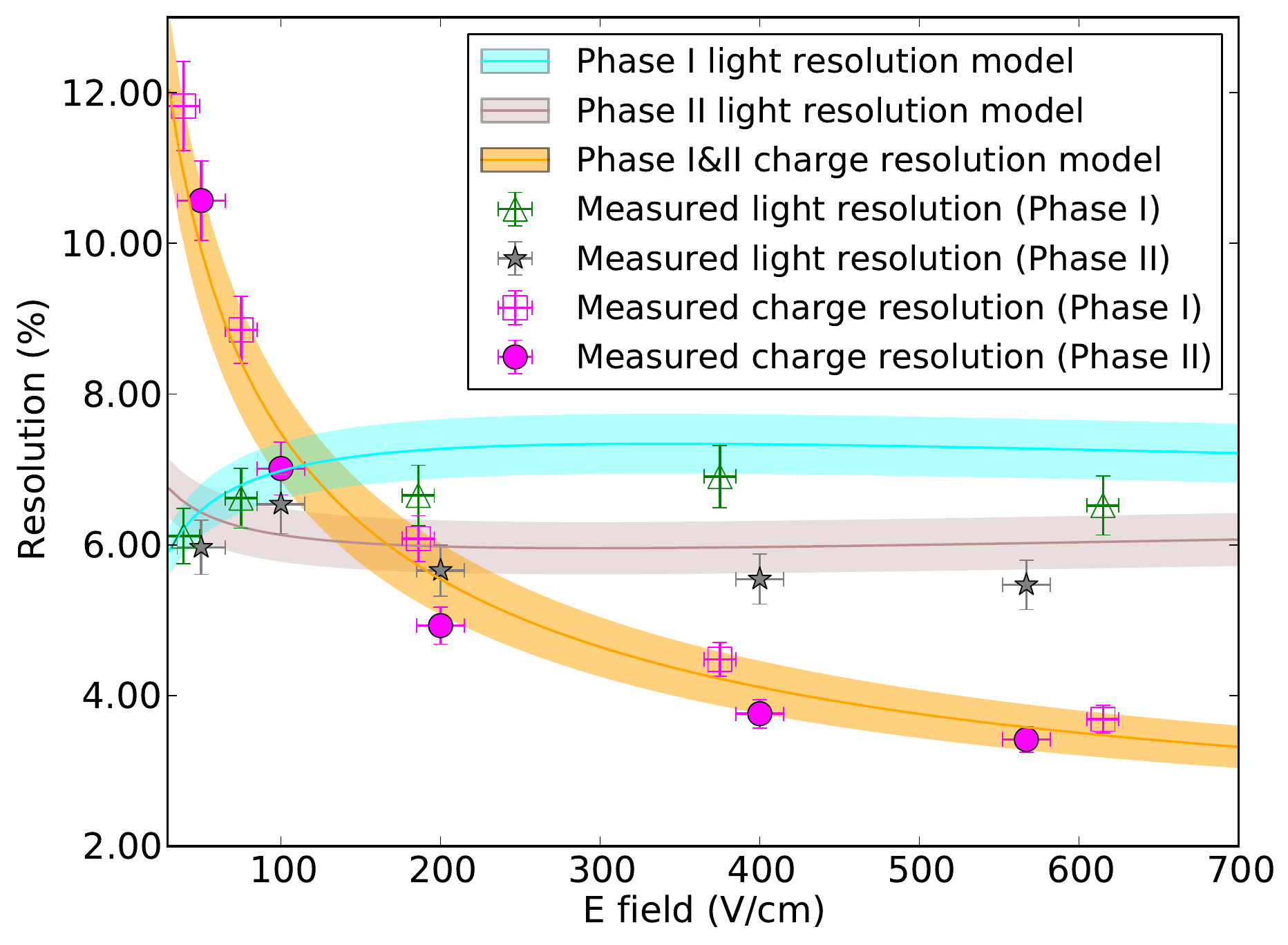}
  \caption{\label{csres}Comparison of measured light and charge energy resolution at 2615~keV under different electric fields with predictions from the resolution model.}
\end{figure}

The resolution for the individual charge and light channels are shown in Fig.~\ref{csres}. The measured values agree with the prediction from the resolution model within the systematic errors. For data taken during Phase~I, the light channel resolution becomes slightly larger as the electric field increases since the APD electronics noise remains constant while the photon yield decreases. In contrast, data taken during Phase~II have lower APD electronics noise compared to Phase~I, and the light channel resolution improves as the recombination fluctuations are reduced at higher electric fields. 

The overall energy resolution is improved after the electronics upgrade, due to the lowered APD noise. In addition, the rotated resolution improves with increasing electric field in both the model and data since more energy is collected by the charge channels, which have relatively lower noise.  The model predicts the resolution at $-12$~kV cathode bias (corresponding to an electric field of 567~V/cm) of $\sigma_E/E = [1.25 \pm 0.08$~(syst.) 
$\pm 0.02$~(stat.)]\%, achieved at the optimal angle in the model of $\theta=39\pm2^{\circ}$. This value is consistent, within error, with the rotation angle of $43\pm3^{\circ}$ measured in Fig.~\ref{gauss2d_Th} under the same field. The model also agrees with the measured Phase~II resolution of 1.23\% for the non-denoised data~\cite{EXOlatest}. As shown in Fig.~\ref{ang}, the model matches the data for both Phase~I and Phase~II (before and after the electronics upgrade) and can reproduce the electric field dependence within systematic errors.

In summary, this resolution model is consistent with the observed resolution in EXO-200 and can be used to predict the performance of future LXe detectors once all relevant quantities are measured. Relative to EXO-200, the energy resolution for LXe detectors can be further improved by eliminating the dominant sources of noise above, e.g. electronics noise in the photo-detector readout~\cite{Albert:2017hjq}.

\subsection{ Measurement of the recombination efficiency $\epsilon_r$\label{intrinsicrho}}
In previous sections, it is assumed that every recombined electron generates a VUV photon, such that the recombination fluctuations of the charge channel $\sigma_{rq}^2$ and the light channel $\sigma_{rp}^2$ are identical. In general, the fluctuation for the light channel is $\sigma_{rp}^2 = \epsilon_r\sigma_{rq}^2$, where $\epsilon_r$ is the recombination efficiency, i.e., the fraction of recombined electrons which produce a VUV photon. The covariance between the charge and light response is then $\mathrm{Cov}_{q,p}=-\epsilon_r\sigma_{rq}^2$. A value of $\epsilon_r = 1$ would correspond to perfect recombination efficiency. If an absolute calibration of the light detection efficiency $\epsilon_p$ were available, $\epsilon_r$ could be directly measured from the change in the charge and light response at different fields, but since the measurement of $\epsilon_p$ above relies on the assumption that $\epsilon_r = 1$, the previous measurements cannot be used directly to determine $\epsilon_r$. However, we can use the agreement of the resolution model with the observed variance in data to test this assumption.

Since both the light channel fluctuations and the covariance term can be written in terms of $\sigma_{rq}$, the variance in the total number of quanta for a general value of $\epsilon_r$ is:  
\begin{equation}
\label{general}
  \sigma_n^2=\cos^2(\theta)\sigma_q^2 + \sin^2(\theta)\sigma_p^2 -2\epsilon_r\cos(\theta)\sin(\theta)\sigma_{rq}^2
\end{equation}
where the total charge and light variances are: $\sigma_{q}^2 = \sigma_{rq}^2 + \sigma_{q,noise}^2$ and $\sigma_p^2 = \epsilon_r^2\sigma_{rq}^2 + \frac{1}{\epsilon_p^2}[n_{PE}(F_N - 1+B^2)+\sigma_{p,noise}^2+n_{PE}^2\sigma_{NU}^2]$. $\sigma_{rq}$ in Eq.~\ref{general} denotes the recombination fluctuations in the electron count and is estimated by subtracting the electronics noise term from the measured total charge variance. 

Using the measurements of the various detector parameters described above (and relaxing the assumption of $\epsilon_r=1$ used to previously estimate $\epsilon_p$), Eq.~\ref{general} is a function of only two unknown parameters, $\epsilon_p$ and $\epsilon_r$, with the optimal rotation angle $\theta$ determined by minimizing the rotated resolution for each set of parameters. We construct a $\chi^2$ statistic by comparing the resolution predicted by the model with the experimental data: $\chi^2=\sum\frac{[X(\epsilon_p,\epsilon_r)-X_{exp}]^2}{\sigma_X^2}$, where the sum is over all measurements performed under various electric fields, $X(\epsilon_p,\epsilon_r)$ is the predicted observable for photon detection efficiency $\epsilon_p$ and recombination efficiency $\epsilon_r$, $X_{exp}$ is the measured value from experimental data, and $\sigma_X$ is the uncertainty of the measured observable.  The values, $X$, used in the fit include the charge resolution, light resolution, rotated resolution, the change in the mean number of electrons, $\Delta n_q$, and the decrease in PE counts, $\Delta n_{PE}$, as the electric field changes.  Since the mean change in the electron and PE counts are related by $\Delta n_{PE}=\Delta n_q\cdot\epsilon_r\cdot\epsilon_p$, a simultaneous fit to these data and the resolution in each channel is used to constrain $\epsilon_p$ and $\epsilon_r$. The fit contains nuisance parameters incorporating possible systematic errors on $B$ and the overall calibration of the number of PE/ADC counts, which are profiled over when calculating the $\chi^2$.

\begin{figure}[t]
\centering 
    \includegraphics[width=0.48\textwidth]{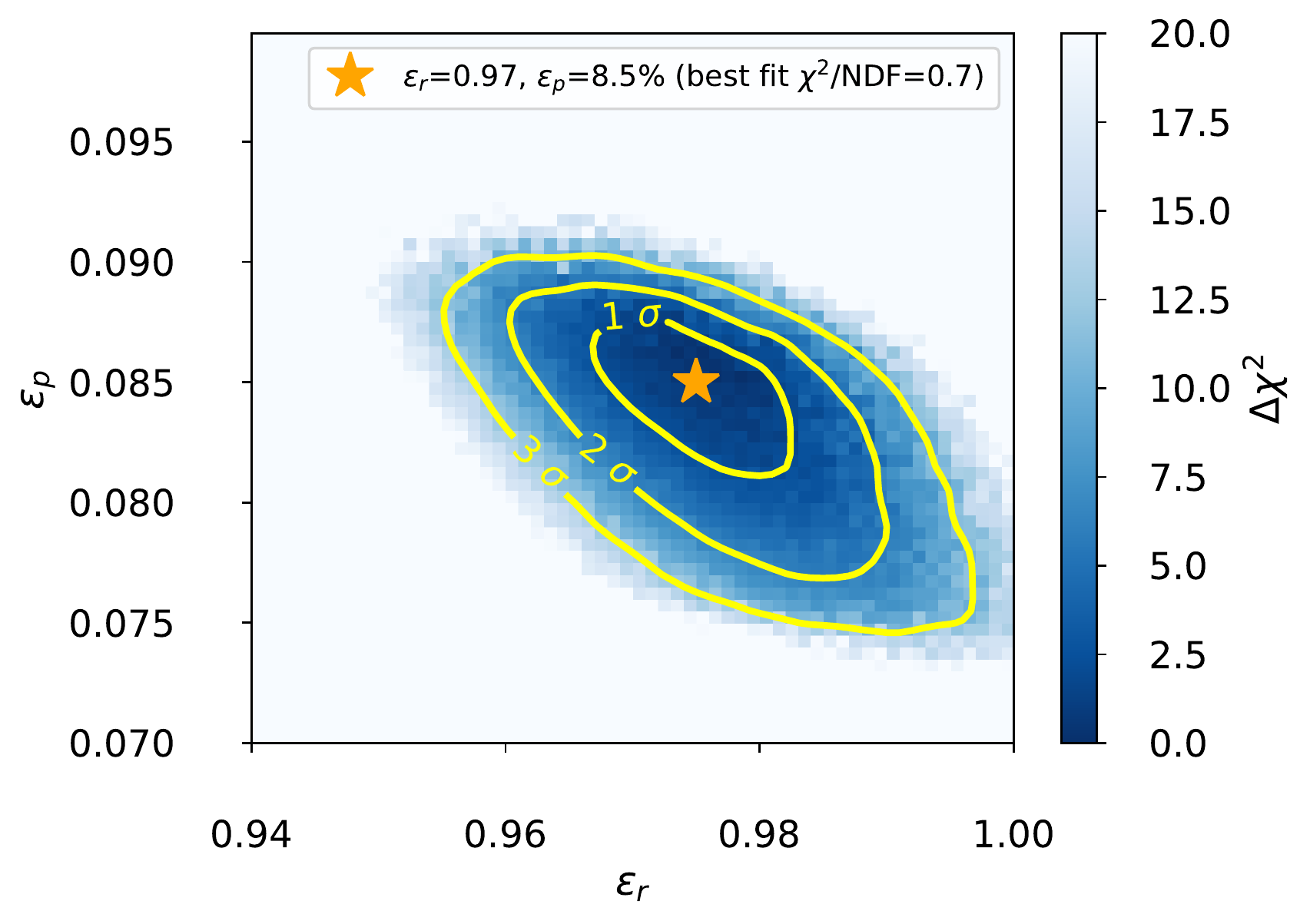}
  \caption{\label{bestre} Change in the $\chi^2$ relative to the best fit point (star) obtained by fitting the predicted energy resolution and the expected change in the light and charge yields versus electric field for various values of the photon detection efficiency, $\epsilon_p$, and intrinsic recombination efficiency, $\epsilon_r$, to the observed data.}
\end{figure}

The combined $\chi^2$ fit is performed using the $^{228}$Th source calibration data from Phase II as described in Sec.~\ref{sec:calibrations}, for which there is a single high-statistics photopeak that can be used to measure the data values at each electric field. The results of the $\chi^2$ fit to these data are shown in Fig.~\ref{bestre}.  The best fit occurs at $\epsilon_r=0.97$ and $\epsilon_p=8.5\%$.  While the fit finds a value of $\epsilon_r$ very close to 1 as assumed in previous sections, it does prefer a non-unity value for this parameter at 3$\sigma$, indicating that the best fit to the resolution in the model occurs if 1\textendash4\% of recombining electron-ion pairs do not produce a detectable photon.  In addition, relaxing the assumption on $\epsilon_r$ does not substantially affect the best-fit value of $\epsilon_p$, and the $W$ value that would be inferred using this best-fit point agrees within systematic errors with that reported in Sec.~\ref{chargelightyield}. These results depend on the accuracy of the semi-empirical resolution model described above, and unknown sources of systematic errors or contributions to the overall resolution that are not included in the model could affect the best fit value for $\epsilon_r$ and its consistency with unity. 
\section{conclusion}
In this paper, we have measured the ionization and scintillation yields in LXe for $\gamma$-rays from $^{228}$Th (2615~keV), $^{226}$Ra (1764~keV) and $^{60}$Co (1332~keV and 1173~keV) calibration sources at a variety of electric fields. These results benefit from the use of a large, single-phase TPC, for which the charge response can be absolutely calibrated.
Using this calibration, this work provides detailed measurements of the absolute yields for $\gamma$s from calibration sources in the 1--2.5~MeV energy range. The measured yields differ by $\sim$10\% in the charge channel and $\sim$20\% in the light channel from the NEST simulation.  The measured $W$-value of $11.5\pm0.5 (syst.)\pm0.1 (stat.)$~eV for MeV scale $\gamma$ interactions differs by $\sim$15\% from the value currently adopted by NEST.  

A semi-empirical model consistent with the energy resolution measured in EXO-200 data at a variety of electric fields is provided, based on direct measurements of the relevant detector parameters, including recombination fluctuations in the number of electrons and photons at various energies. This model can account for the dominant sources of noise in the EXO-200 energy measurement, and places constraints on the recombination efficiency of electron-ion pairs in LXe.  

A number of systematic cross-checks have been performed on these measurements, using the detailed understanding of the EXO-200 detector response developed throughout the operation of the experiment. These measurements take advantage of a large detector with a well-understood energy response, which is based on a comprehensive detector Monte Carlo simulation. These results provide new measurements of the absolute yields of charge and light in LXe at MeV energies, extending previous measurements primarily performed with smaller R\&D systems at lower energies. The measurements presented here can also guide simulations of the charge and light production in future $0\nu\beta\beta$ and rare event searches employing LXe.  
\begin{acknowledgments}
We would like to thank Matthew Szydagis and Jason Brodsky for helpful discussions related to NEST. EXO-200 is supported by DOE and NSF in the United States, NSERC in Canada, SNF in Switzerland, IBS in Korea, RFBR (18-02-00550) in Russia, DFG in Germany, and CAS and ISTCP in China. EXO-200 data analysis and simulation uses resources of the National Energy Research Scientific Computing Center (NERSC). We gratefully acknowledge the KARMEN collaboration for supplying the cosmic-ray veto detectors, and the WIPP for their hospitality.
\end{acknowledgments}
\appendix
\section{\label{APDnon}APD gain non-uniformity}
Each APD channel consists of $\sim$7 individual APDs ganged together in a single readout channel~\cite{Auger}, which are biased with a single voltage for the entire gang. The gain non-uniformity among the APDs within a gang may cause variation in the measured photon count for a given photopeak. 

We make use of the lightmap described in Sec.~\ref{res1}, which provides an empirical measurement of the PE number created in each APD channel given a scintillation cluster's 3D position, to determine this additional $\sigma_{NU}$ in Eq.~\ref{sigmap2}. The APD plane in TPC1 (TPC2) is located at $z = 204$~mm ($z = -204$~mm). To select energy deposits occurring near the APD plane, but within the fiducial volume of the detector, we select clusters occurring at $z = 182\pm1$~mm ($z = -182\pm1$~mm), and record the number of PEs collected by the APD closest to each event. For such events, a larger number of photons are collected by the single APD in each readout gang closest to the event position, which allows the estimation of the gain variation among the APDs within the gang. During EXO-200 operations, five APD gangs could not be operated due to hardware problems and were not considered in the measurement.

\begin{figure*}
\begin{minipage}[t]{0.45\linewidth}
\includegraphics[width=1\linewidth]{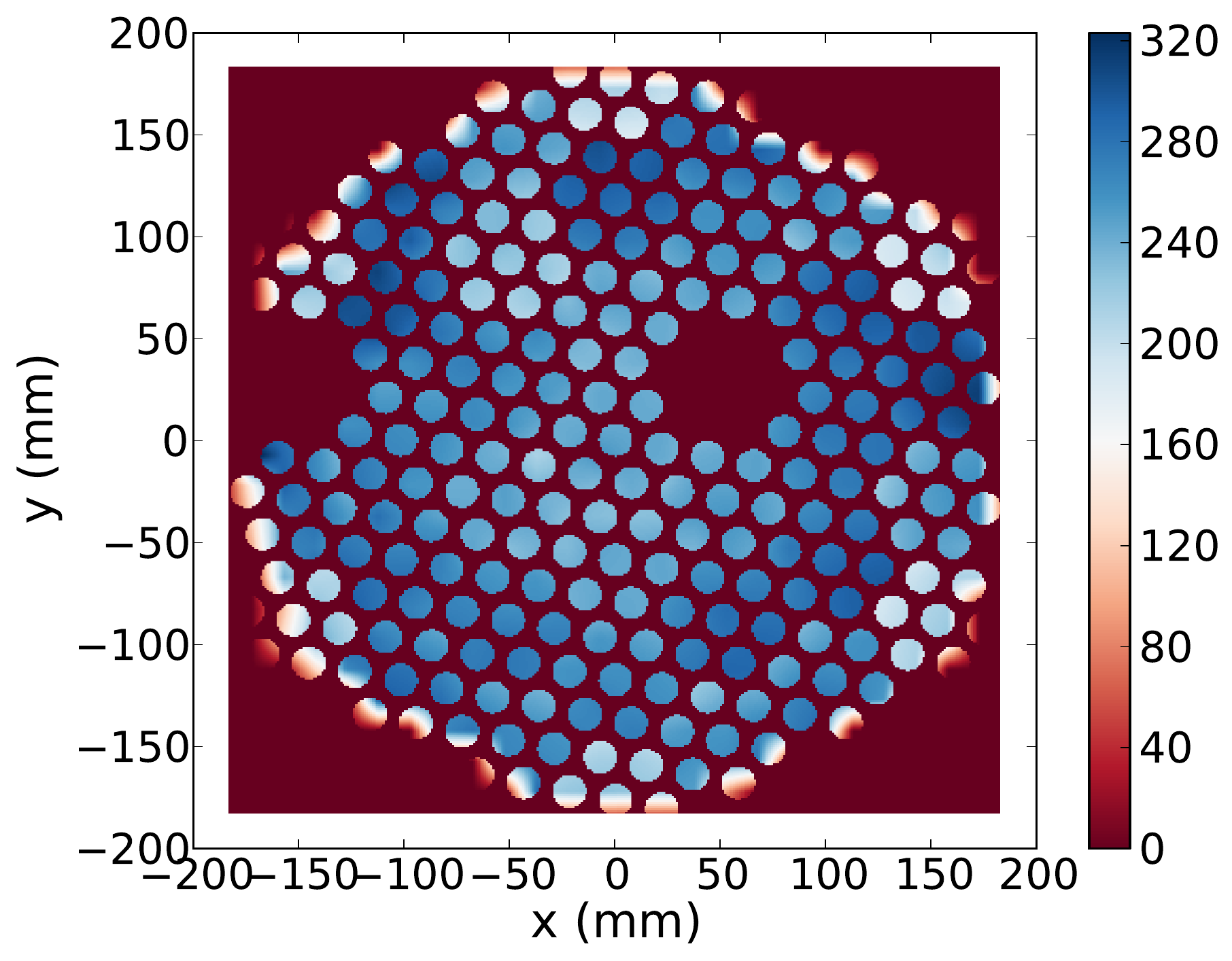}
\newline(a)
\end{minipage}\hfill
\begin{minipage}[t]{0.45\linewidth}
\includegraphics[width=1\linewidth]{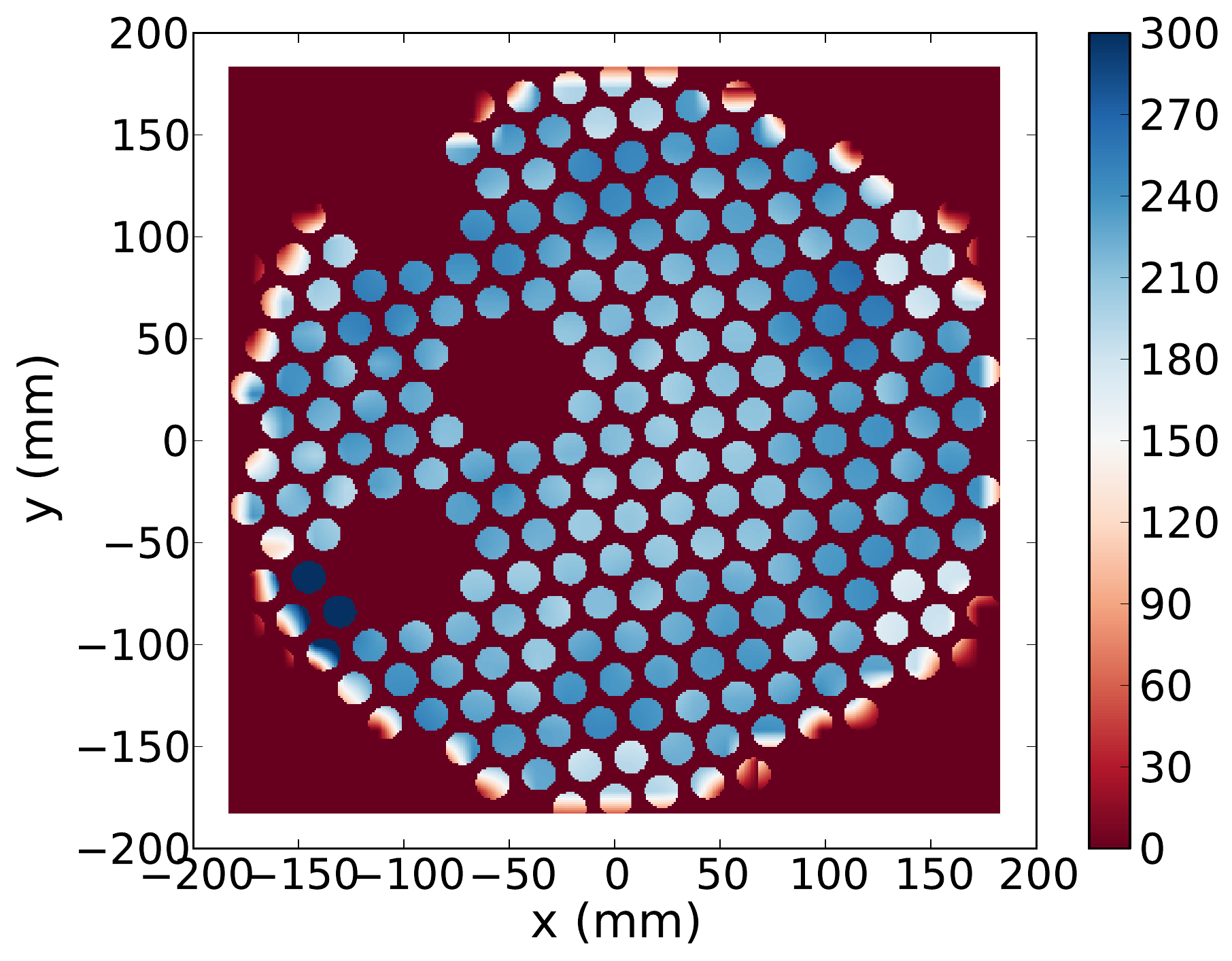}
\newline(b)
\end{minipage}
\caption{PE counts on each APD channel from a scintillation cluster right above it at (a) $z$~=~50~mm for TPC1 and (b) $z$~=~-50~mm for TPC2 measured using the lightmap data.}\label{nonuni20}
\end{figure*}
  
The gain non-uniformity---namely the differences in gain at fixed bias for APDs within the same channel---is measured by first calculating the variation among the total PE number for the 7 component APDs and dividing by the mean number of PEs. The average APD gain non-uniformity is measured to be 2.4\%$\pm$1.1\%, in which the error denotes the spread in the non-uniformity values among the measurements on different gangs. Edge channels with APDs outside the fiducial volume are not included in the average.

For events occurring very near the APD plane, the measured gain non-uniformity can have an impact on the rotated resolution that is non-negligible compared to other terms. However, the non-uniformity has a substantially smaller effect on $\sigma_p^2$ when the event is far away from the anode, as can be seen from Fig.~\ref{nonuni20}, where the response of each channel to a scintillation cluster that is $\sim$150~mm directly above the APD plane is shown. Only $x$-$y$ positions of clusters directly above the circular face of an APD are considered. In this case, the uniform distribution of photons across each gang due to the smaller solid angle variation with position and relatively larger amount of reflected light smooths out the effects of gain non-uniformity within each gang. For the source calibrations considered here, the sources were positioned near the cathode, at maximal distance from each APD plane.  After accounting for the uniformity of the response on each gang seen in Fig.~\ref{nonuni20}, the $\sigma_{NU}$ term is estimated to be $<$0.2\% and can be neglected from the calculation of $\sigma_p^2$ for these results.

\bibliography{apssamp}
\end{document}